\documentclass{elsart}
\pdfoutput=1

\usepackage{amsmath}
\usepackage{amssymb}
\usepackage{color}
\usepackage{url}

\usepackage{algorithm}
\usepackage[noend]{algorithmic} 
\usepackage{verbatim}
 \newtheorem{theorem}{Theorem}
 \newtheorem{proposition}{Proposition}

 \newtheorem{corollary}{Corollary}
 
\newtheorem{lemma}{Lemma}
\usepackage{subfig}

\usepackage{graphicx}

\graphicspath{{../}{../plots/}{../../results/001/}{../../results/002/}{../../results/003/}{../../../svn/lemurpython/timeseries/}}

\usepackage{setspace}

\begin{document}

\begin{frontmatter}



\title{Faster Retrieval with a Two-Pass  Dynamic-Time-Warping  Lower Bound}



 \author{Daniel Lemire\thanksref{label2}}
 \ead{lemire@acm.org}
 \thanks[label2]{phone: 00+1+514 987-3000 ext. 2835, fax: 00+1+514 843-2160}
 \address{LICEF, Universit\'e du Qu\'ebec \`a Montr\'eal (UQAM), 100 Sherbrooke West, Montreal~(Quebec), H2X 3P2 Canada
}


\begin{abstract}
The Dynamic Time Warping (DTW) is a popular similarity measure between time series.
The DTW fails to satisfy the triangle inequality and its computation 
requires quadratic time.
Hence, to find  closest neighbors quickly, we use bounding 
techniques.
We can avoid most DTW computations with an inexpensive lower bound (LB\_Keogh).
We compare LB\_Keogh with a tighter lower bound (LB\_Improved). 
We find that LB\_Improved-based search is faster.
 As an example, 
our approach is 2--3~times faster over random-walk and shape time series.
%
%
\end{abstract}

\begin{keyword}
time series \sep very large databases \sep indexing \sep classification

\end{keyword}
\end{frontmatter}

\section{Introduction}

\begin{figure}[hb]
\centering\includegraphics[width=0.5\columnwidth]{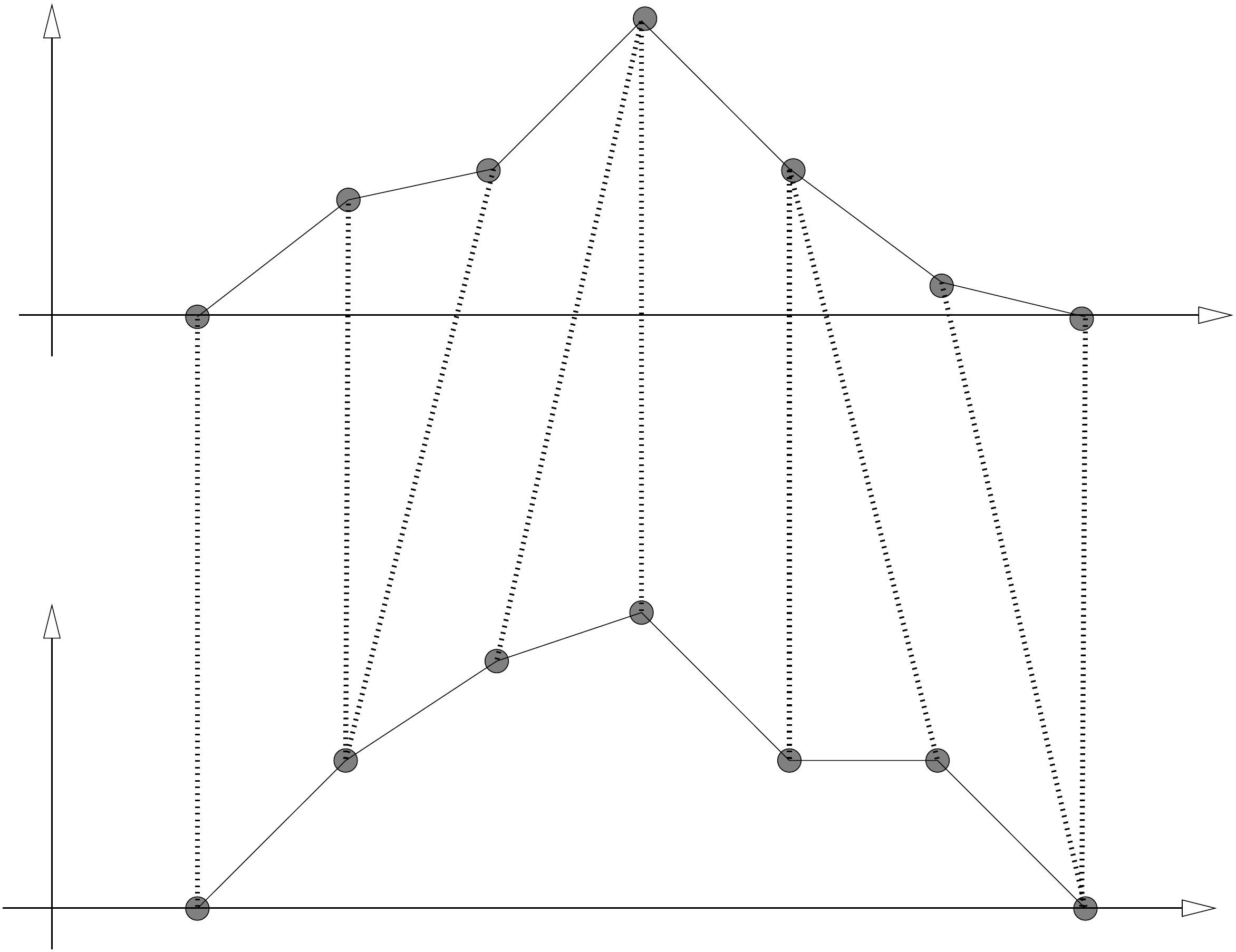}
\caption{\label{fig:example}Dynamic Time Warping example}
\end{figure}

Dynamic Time Warping (DTW) was initially introduced
to recognize spoken words~\cite{sakoe1978dpa}, but
it has since been applied to a wide range of information
retrieval and database problems:
handwriting recognition~\cite{bahlmann2004wio,niels2005udt},
 signature recognition~\cite{1219544,chang2007mdt}, image de-interlacing~\cite{almog2005},
  appearance matching for security purposes~\cite{Kale2004}, 
  whale vocalization classification~\cite{brown2006cvk}, 
  query by humming~\cite{4432642,zhu2003wie},
  classification of motor activities~\cite{muscillo2007cma},
 face localization~\cite{lopez2007flc},
 chromosome classification~\cite{legrand2007ccu},
  shape retrieval~\cite{TPAMI.2005.21,marzal2006cbs}, and so on. 
Unlike the Euclidean distance, DTW optimally aligns or ``warps'' 
the data points of
two time series (see Fig.~\ref{fig:example}). 

When the distance between two time series forms a metric, such as
the Euclidean distance or the Hamming distance,
several indexing or search techniques have been 
proposed~\cite{502808,1221301,958948,230528,1221193}.
However, even assuming that we have a metric, Weber et al. have shown
that the performance of any indexing scheme degrades  to that of a sequential scan,
when there are more than a few dimensions~\cite{671192}.
Otherwise---when the distance is not a metric or that
the number of dimensions is too large---we use 
bounding techniques such as the
Generic multimedia object indexing (GEMINI)~\cite{faloutsos1996smd}.
We  quickly discard (most) false positives
by computing a lower bound. 

Ratanamahatana and Keogh~\cite{ratanamahatana2005tmd} argue that their lower bound
(LB\_Keogh) cannot be improved upon. To make their point, they report that LB\_Keogh
allows them to prune out over 90\% of all
DTW computations on several data sets. 

We are able to improve upon LB\_Keogh as follows.
The first step of our two-pass approach is LB\_Keogh itself.
If this first lower bound is sufficient to discard the candidate, then
the computation terminates and the next candidate is considered.
Otherwise, we process the time series a second time to increase the
lower bound (see Fig.~\ref{fancyplot}). If this second lower bound is large enough,
the candidate is pruned, otherwise we compute the full DTW\@. 
We show experimentally
that the two-pass approach can be several times faster.

The paper is organized as follows. In Section~\ref{sec:dtw}, we 
 define the DTW in a generic manner
as the minimization of the $l_p$ norm ($\text{DTW}_p$).
Among other things, we show that if $x$ and $y$ are separated by a constant ($x\geq c \geq y$ or $x\leq c \leq y$)
then the $\text{DTW}_1$ is the $l_1$ norm (see Proposition~\ref{prof:band1}).
In Section~\ref{sec:definingUL}, we compute generic lower bounds on the DTW and their
approximation errors using warping envelopes.
In Section~\ref{sec:envelopes}, we show how to compute the warping
envelopes quickly. 
The next two sections introduce LB\_Keogh and LB\_Improved respectively.
Section~\ref{sec:multiindex} presents the application of these lower bounds
for multidimensional indexing whereas the last section presents an experimental comparison.

\section{Conventions}

Time series are arrays of values measured at certain times.
For simplicity, we assume a regular
 sampling rate so that time series are generic arrays of floating-point values.
Time series have length $n$ and
are indexed from 1 to $n$.
The $l_p$ norm of  $x$ 
is $\Vert x \Vert_p= (\sum_i \vert x_i\vert ^p)^{1/p}$ for any integer
$0<p<\infty$ and $\Vert x \Vert_\infty= \max_i\vert  x_i\vert$. 
The $l_p$ distance between $x$ and $y$ is  $\Vert x - y\Vert_p$
and it satisfies
the triangle inequality $\Vert x - z\Vert_p\leq \Vert x - y\Vert_p+ \Vert y - z\Vert_p$ for $1\leq p\leq \infty$.
The distance between a point $x$ and a set or region $S$ is $d(x,S)= \min_{y\in S} d(x,y)$.
Other conventions are summarized in Table~\ref{tab:commonnot}.

\begin{table}[bth]
\caption{\label{tab:commonnot}Frequently used conventions}
\begin{tabular}{ccc}\hline
 $n$ & length of a time series \\
$\Vert x \Vert_p$ & $l_p$ norm \\
$\text{DTW}_{p}$ & monotonic DTW \\
$\text{NDTW}_{p}$ & non-monotonic DTW \\
$w$ & DTW locality constraint\\ 
$U(x), L(x)$ & warping envelope (see Section~\ref{sec:definingUL})\\
$H(x,y)$ & projection of $x$ on $y$ (see Equation~\ref{eqn:hxy})\\
\hline
\end{tabular}
\end{table}
\section{Related Works}

Beside DTW, several similarity metrics have been
proposed including the
directed and general  Hausdorff distance, Pearson's correlation,
 nonlinear elastic matching distance~\cite{veltkamp2001sms}, 
 Edit distance with Real Penalty (ERP)~\cite{chen:mln},
Needleman-Wunsch similarity~\cite{needleman1970gma},
Smith-Waterman similarity~\cite{smith1981icm},
and SimilB~\cite{1340465}.

Boundary-based lower-bound functions sometimes outperform LB\_Keogh~\cite{zhou2007bbl}.
We can also quantize~\cite{1166502} 
the time series.

Sakurai et al.~\cite{Sakurai2005} have shown that
retrieval under the DTW can be faster by mixing progressively finer resolution
and by applying early abandoning~\cite{1106385}
to the dynamic programming computation.


\section{Dynamic Time Warping}
\label{sec:dtw}
A many-to-many matching between the data points in time series $x$ and the data point
in time series $y$ matches every data point $x_i$ in $x$ with at least one data point $y_j$ in $y$,
and every data point in $y$ with at least a data point in $x$. The set of 
matches $(i,j)$ forms a \emph{warping path} $\Gamma$.
We define the DTW as the minimization of the $l_p$ norm of the differences $\{ x_i-y_j  \}_{(i,j)\in \Gamma}$
over all warping paths. 
A warping path is minimal if there is no subset $\Gamma'$ of $\Gamma$ 
forming an warping path: for simplicity we require all warping paths to be minimal.

In computing the DTW distance, we commonly require the warping to remain
local. For time series $x$ and $y$, we align values 
$x_i$ and $y_j$ only if $\vert i- j \vert \leq  w$ for some locality constraint $w\geq 0 $~\cite{sakoe1978dpa}.
When $w=0$, the DTW becomes the $l_p$ distance whereas when $w\geq n$, the DTW has no locality constraint.
The value of the DTW diminishes monotonically as $w$ increases.
(We do not consider other forms of locality constraints such as the Itakura parallelogram~\cite{itakura1975mpr}.)
 
 Other than locality, DTW can be monotonic: if we align value $x_i$
 with value $y_j$, then we cannot align value $x_{i+1}$ with a value
 appearing before $y_j$ ($y_{j'}$ for $j'<j$).

We note the DTW distance between $x$ and $y$ using the $l_p$ norm as $\text{DTW}_p(x,y)$
when it is monotonic and as $\text{NDTW}_p(x,y)$ when monotonicity is not required.

By dynamic programming, the monotonic DTW  requires $O(w n)$ time. A typical value of $w$ is $n/10$~\cite{ratanamahatana2005tmd} 
so that the DTW is in~$O(n^2)$.
To compute the DTW, we use the following recursive formula.
Given an array $x$, we write the suffix starting at position $i$, $x_{(i)}= x_i,x_{i+1},\ldots,x_n$.
The symbol $\oplus$ is the exclusive or.  Write $q_{i,j}= \text{DTW}_p(x_{(i)},y_{(j)})^p$ so that $\text{DTW}_p(x,y)=\sqrt[p]{q_{1,1}}$,
then
\begin{align*}
q_{i,j}= \begin{cases}0  & \text{if $\vert x_{(i)}\vert = \vert (y_{(j)}\vert =0$}\\
\infty  & \begin{matrix}\text{if $\vert x_{(i)}\vert = 0 \oplus \vert y_{(j)}\vert =0$}\\
\text{or  $\vert i- j \vert > w$ }\end{matrix}\\
 \begin{matrix}\vert x_i-y_j \vert^p +\\ \min (
 q_{i+1,j}, q_{i,j+1}, q_{i+1,j+1})\end{matrix} & \text{otherwise.}
\end{cases}
\end{align*}
For $p=\infty$, we rewrite
the preceding recursive formula with 
$q_{i,j}= \text{DTW}_\infty(x_{(i)},y_{(j)})$, and $q_{i,j}=\max (\vert x_i-y_j \vert, \min (
 q_{i+1,j}, q_{i,j+1}, q_{i+1,j+1}))$ when $\vert x_{(i)}\vert \neq 0$,  $\vert y_{(j)}\vert \neq 0$, and $\vert i-j\vert \leq w$.

We can compute $\text{NDTW}_1$ 
without locality constraint in $O(n \log n)$~\cite{1275562}: if the values of the
time series are already sorted, the computation is in $O(n)$ time.

We can express the solution of the DTW problem  
 as an alignment of the two initial time series (such as $x=0,1,1,0$ and $y=0,1,0,0$) where some of the 
values are repeated (such as $x'=0,1,1,0,\textbf{0}$ and $y'=0,1,\textbf{1},0,0$).
If we allow non-monotonicity (NDTW), then values can also be inverted.

The non-monotonic DTW is no larger than the monotonic DTW which is itself no 
larger than the $l_p$ norm: $\text{NDTW}_p(x,y) \leq \text{DTW}_p(x,y) \leq \Vert x-y\Vert_p$ for all $0<p\leq \infty$.


The $\text{DTW}_1$ has the property that if 
the time series are value-separated, then the DTW is the $l_1$ norm
as the next proposition shows.
 In  Figs.~\ref{fig:pykeogh-2} and \ref{fig:pykeogh-3},
we present value-separated functions: their $\text{DTW}_1$ is the area between the curves.

\begin{proposition}\label{prof:band1}If $x$ and $y$
are such that  either $x\geq c \geq y$ or $x\leq c \leq y$ for some constant $c$,
then $\text{DTW}_1(x,y)=\text{NDTW}_1(x,y)= \Vert x-y \Vert_1$.
\end{proposition}
\begin{pf}Assume $x\geq c \geq y$.
Consider the two aligned (and extended) time series $x', y'$
 such that  $\text{NDTW}_1(x,y)=\Vert x'-y' \Vert_1$. We have that
$x'\geq c \geq y'$ and $\text{NDTW}_1(x,y)= \Vert x'-y' \Vert_1 =\sum_i \vert x'_i - y'_i\vert
= \sum_i \vert x'_i - c\vert  + \vert c- y'_i\vert = \Vert x'-c \Vert_1 + \Vert c-y' \Vert_1 \geq
\Vert x-c \Vert_1 + \Vert c-y \Vert_1 =  \Vert x-y \Vert_1$. Since 
we also have  $\text{NDTW}_1(x,y) \leq \text{DTW}_1(x,y) \leq \Vert x-y\Vert_1$, the equality follows.
\end{pf}

Proposition~\ref{prof:band1} does not hold for $p>1$: $\text{DTW}_2((0,0,1,0), (2,3,2,2))=\sqrt{17}$
whereas $\Vert(0,0,1,0) - (2,3,2,2)\Vert_2=\sqrt{18}$.

\section{Computing Lower Bounds on the DTW}

\label{sec:definingUL}
\begin{figure}
\centering
\includegraphics[width=0.7\columnwidth]{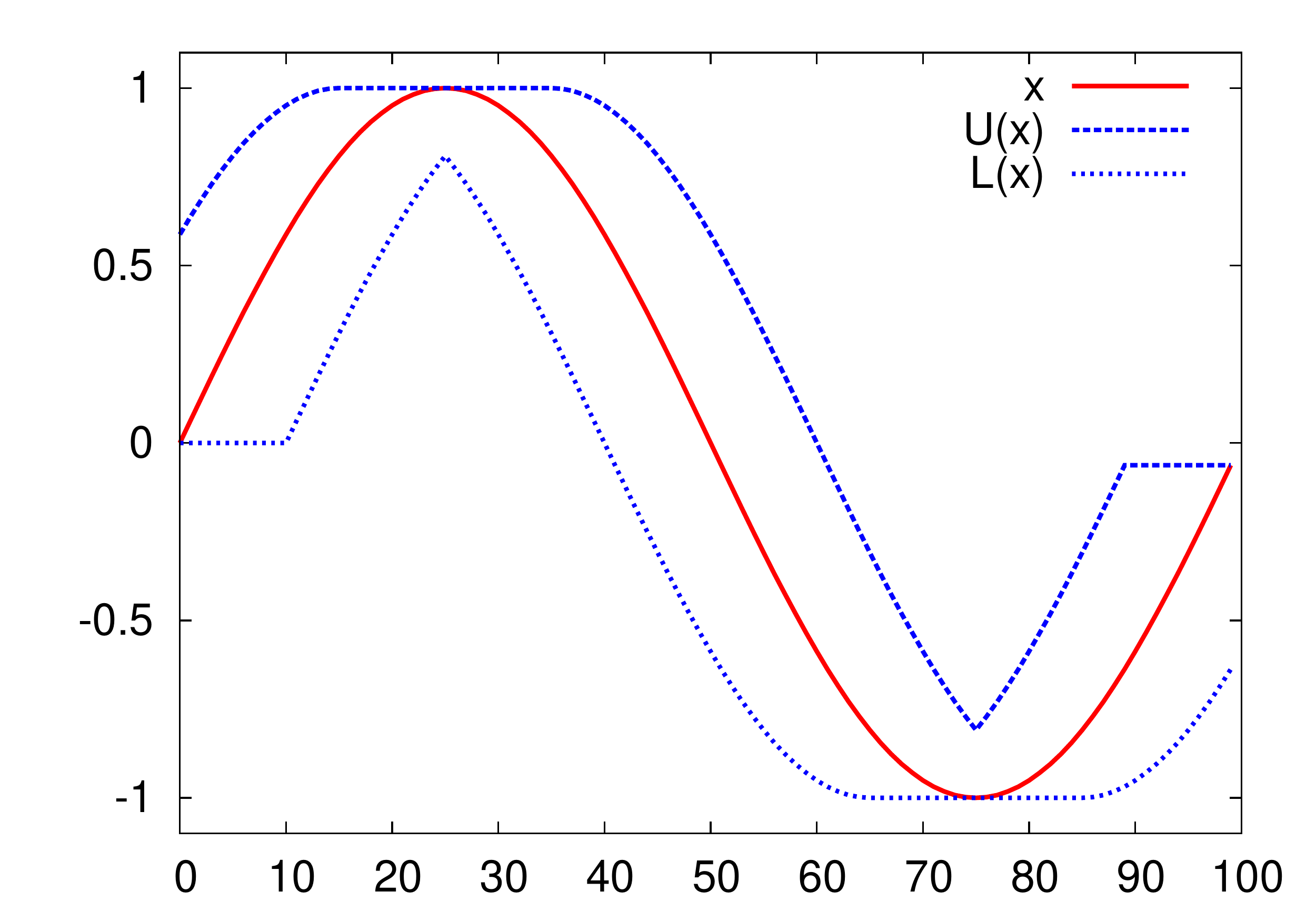}
\caption{\label{fig:pykeogh-4}Warping envelope example}
\end{figure}

Given a time series $x$, define $U(x)_i = \max_k \{x_k |\, \vert k-i \vert \leq w \}$
and $L(x)_i = \min_k \{x_k |\, \vert k-i \vert \leq w \}$ for $i=1, \ldots, n$. The pair $U(x)$ and $L(x)$ forms
the warping envelope of $x$ (see Fig.~\ref{fig:pykeogh-4}). We leave the locality constraint $w$ implicit.

The theorem of this section has an elementary proof  requiring only
the following technical lemma.

\begin{lemma}\label{lemma:technical}If $b\in[a,c]$ then $(c-a)^p \geq (c-b)^p+(b-a)^p$ for $1\leq p< \infty$.
\end{lemma}
\begin{pf}For $p=1$, $(c-b)^p+(b-a)^p=(c-a)^p$. For $p>1$, by deriving
  $(c-b)^p+(b-a)^p$ with respect to $b$, we can show that it is minimized when 
$b=(c+a)/2$ and maximized when $b\in \{a,c\}$. The maximal value is $(c-a)^p$. Hence the result.
\end{pf}

The following theorem introduces a generic result that we  use to derive two
lower bounds for the DTW including the original Keogh-Ratanamahatana result~\cite{keogh2005eid}.
Indeed, this new result not only implies the lower bound LB\_Keogh,
but it also provides a lower bound to  the error made by LB\_Keogh, thus allowing a
tighter lower bound (LB\_Improved).
%

\begin{theorem}\label{thm:mainthm}
Given two equal-length time series $x$ and $y$ and $1\leq p<\infty$, then for any time series
$h$ satisfying $x_i\geq  h_i\geq U(y)_i$ or  $x_i \leq h_i \leq L(y)_i$
or $h_i=x_i$ for all indexes $i$, we have
\begin{align*}
\text{DTW}_p(x,y)^p & \geq \text{NDTW}_p(x,y)^p \\
                    & \geq   \Vert x-h \Vert_p^p + \text{NDTW}_p(h,y)^p.
\end{align*}
For $p=\infty$, a similar result is true: $\text{DTW}_{\infty}(x,y)  \geq \text{NDTW}_{\infty}(x,y)  \geq   \max(\Vert x-h \Vert_{\infty} ,\text{NDTW}_{\infty}(h,y))$.
\end{theorem}
\begin{pf}
Suppose that $1\leq p<\infty$.
Let $\Gamma$ be a warping path such that $\text{NDTW}_p(x,y)^p = \sum_{(i,j) \in \Gamma} \vert x_i-y_j \vert_p^p $.
By the constraint on $h$ and Lemma~\ref{lemma:technical}, we have that $\vert x_i-y_j \vert^p  \geq \vert x_i-h_i \vert^p  + \vert h_i - y_j \vert^p $
for any $(i,j) \in \Gamma$ since $h_i \in [\min(x_i, y_j), \max(x_i,y_j)]$. 
Hence, we have that  $\text{NDTW}_p(x,y)^p \geq \sum_{(i,j) \in \Gamma}  \vert x_i-h_i \vert^p  + \vert h_i - y_j \vert^p
\geq  \Vert x-h \Vert_p^p  + \sum_{(i,j) \in \Gamma} \vert h_i - y_j \vert^p$.
This proves the result since  $ \sum_{(i,j) \in \Gamma} \vert h_i - y_j \vert \geq  \text{NDTW}_p(h,y)$.
For $p=\infty$, we have that 
\begin{align*}\text{NDTW}_{\infty}(x,y) &=\max_{(i,j)\in \Gamma} \vert x_i-y_j \vert\\
& \leq 
\max_{(i,j)\in \Gamma} \max(\vert x_i-h_i \vert,\vert h_i-y_j \vert)\\
&=\max(\Vert x-h \Vert_{\infty} ,\text{NDTW}_{\infty}(h,y)),
\end{align*}
concluding the proof.
\end{pf}

While Theorem~\ref{thm:mainthm} defines a lower bound ($\Vert x-h \Vert_p$), the next proposition shows that this
lower bound must be a tight approximation as long as $h$ is close to $y$ in the $l_p$ norm.
\begin{proposition}\label{prop:error}
Given two equal-length time series $x$ and $y$, and $1\leq p\leq \infty$ with $h$ as in Theorem~\ref{thm:mainthm},
we have that $\Vert x-h \Vert_p$ approximates both $ \text{DTW}_p(x,y)$
and $ \text{NDTW}_p(x,y)$
within $\Vert h-y \Vert_p$.
\end{proposition}
\begin{pf}
By the triangle inequality over $l_p$, we have  $\Vert x-h \Vert_p +   \Vert h-y \Vert_p \geq\Vert x-y \Vert_p $. 
Since $\Vert x-y \Vert_p \geq  \text{DTW}_p(x,y)$,
we have
$\Vert x-h \Vert_p +   \Vert h-y \Vert_p \geq \text{DTW}_p(x,y)$,
and hence
$ \Vert h-y \Vert_p \geq \text{DTW}_p(x,y) - \Vert x-h \Vert_p $.
This proves the result since by
 Theorem~\ref{thm:mainthm},
we have that $\text{DTW}_p(x,y)\geq \text{NDTW}_p(x,y) \geq \Vert x-h \Vert_p$.
\end{pf}

This bound on the approximation error is reasonably tight. If $x$ and $y$ 
are separated by a constant, then $\text{DTW}_1(x,y) =\Vert x-y \Vert_1$
by Proposition~\ref{prof:band1} and $\Vert x-y \Vert_1 = \sum_i \vert x_i-y_i\vert
=  \sum_i \vert x_i-h_i\vert+\vert h_i-y_i\vert=\Vert x-h \Vert_1+ \Vert h-y \Vert_1 $.
Hence, the approximation error is exactly $\Vert h-y \Vert_1 $ in such instances.

%
%

\section{Warping Envelopes}
\label{sec:envelopes}
The computation of the warping envelope $U(x),L(x)$ requires $O(nw)$ time using the naive approach
of repeatedly computing the maximum and the minimum over windows. 
Instead, we compute the envelope with at most $3n$~comparisons between data-point values~\cite{lemiremaxmin}
using Algorithm~\ref{algo:mystreamingalo}.

\begin{algorithm}
\begin{small}
 \begin{algorithmic}
\STATE \textbf{input} a time series $y$ indexed from $1$ to $n$
\STATE \textbf{input} some DTW locality constraint $w$
\RETURN warping envelope $U,L$ (two time series of length $n$)
\STATE $u$, $l$ $\leftarrow$ empty double-ended queues, we append to ``back'' 
\STATE append $1$ to $u$ and $l$ 
\FOR{ $i$ in $\{2,\ldots,n\}$}\label{alg:mainloop}
\IF{$i\geq w+1$}
\STATE $U_{i-w} \leftarrow y_{\textrm{front}(u)}$, $L_{i-w} \leftarrow y_{\textrm{front}(l)}$ 
\ENDIF
\IF{$y_i > y_{i-1}$}\label{alg:firstcompare}
\STATE pop $u$ from back\label{alg:removemax}
\WHILE{ $y_i > y_{\textrm{back}(u)}$}\label{alg:while1}
\STATE pop $u$ from back 
\ENDWHILE
\ELSE
\STATE pop $l$ from back \label{alg:removemin}
\WHILE{ $y_i < y_{\textrm{back}(l)}$} \label{alg:while2}
\STATE pop $l$ from back
\ENDWHILE
\ENDIF
\STATE append $i$ to $u$ and $l$\label{alg:append}
\IF{$i=2w+1+\textrm{front}(u)$}  \label{alg:ensurescontainted}
\STATE pop $u$ from front
\ELSIF{$i=2w+1+\textrm{front}(l)$}
\STATE pop $l$ from front
\ENDIF
\ENDFOR
\FOR{ $i$ in $\{n+1,\ldots,n+w\}$}\label{alg:secondloop}
\STATE $U_{i-w} \leftarrow y_{\textrm{front}(u)}$, $L_{i-w} \leftarrow y_{\textrm{front}(l)}$
\IF{i-front($u$)$\geq 2w+1$}
\STATE pop $u$ from front
\ENDIF
\IF{i-front($l$)$\geq 2w+1$}
\STATE pop $l$ from front
\ENDIF
\ENDFOR
 \end{algorithmic}
\end{small}
\caption{\label{algo:mystreamingalo}Streaming algorithm to compute the warping envelope using no more than $3n$~comparisons
}
\end{algorithm}

\section{LB\_Keogh}
\label{sec:lbkeogh}
\begin{figure}
\centering
\includegraphics[width=0.7\columnwidth]{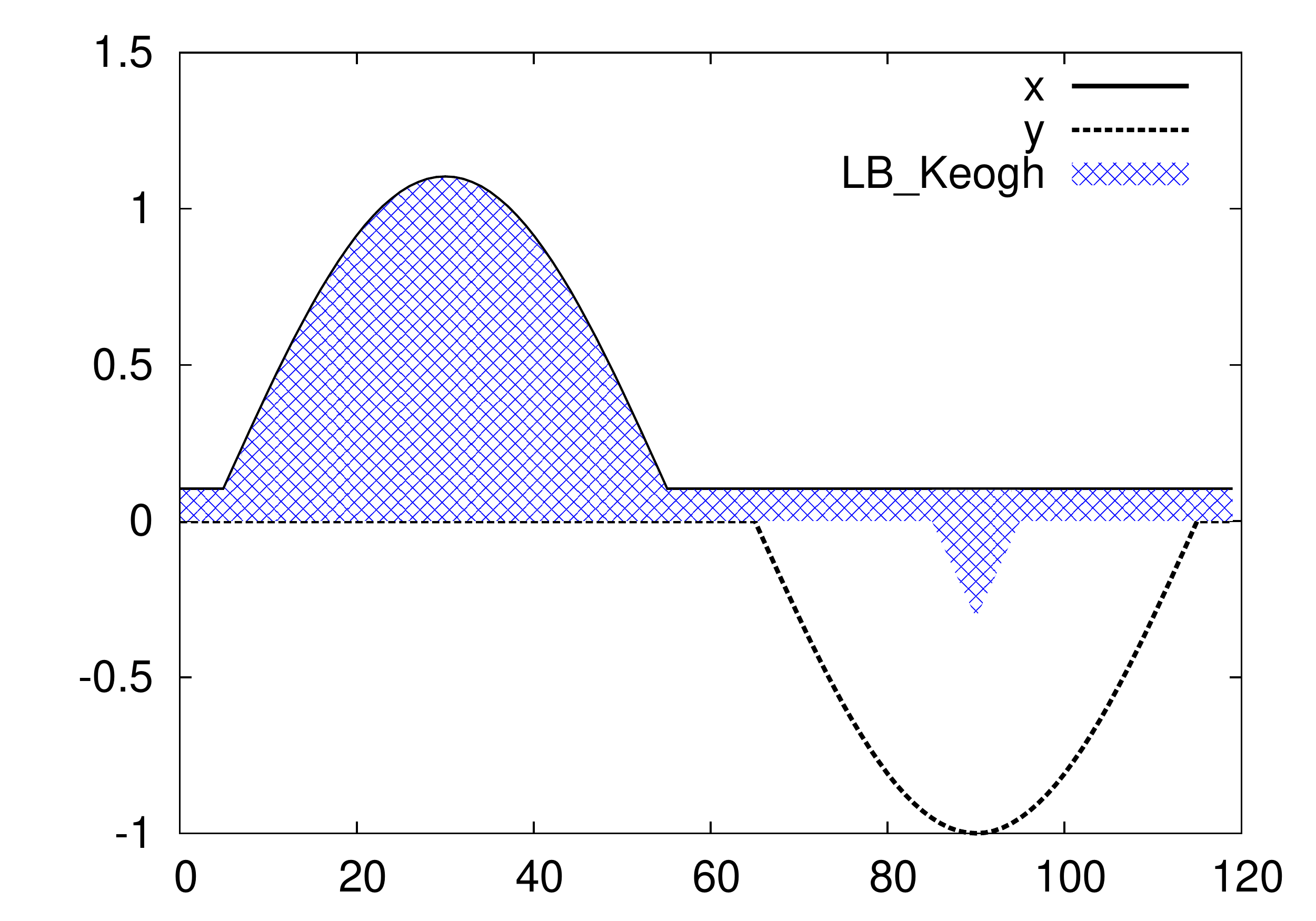}
\caption{\label{fig:pykeogh-2}LB\_Keogh example: the area of the marked region is $\text{LB\_Keogh}_1(x,y)$}
\end{figure}

Let $H(x,y)$ be the \emph{projection of $x$ on $y$} defined as
\begin{align}\label{eqn:hxy}
H(x,y)_{i}= \begin{cases}U(y)_i  & \text{if $x_i\geq U(y)_i$}\\
L(y)_i  & \text{if $x_i\leq L(y)_i$} \\
x_i & \text{otherwise,}
\end{cases}
\end{align}
for $i=1,2,\ldots,n$.
We have that $H(x,y)$ is in the envelope of $y$. By Theorem~\ref{thm:mainthm} and setting $h= H(x,y)$, 
we have that $\text{NDTW}_p(x,y)^p\geq \Vert x- H(x,y)\Vert_p^p + \text{NDTW}_p(H(x,y),y)^p$ for $1 \leq p<\infty$.
Write $\text{LB\_Keogh}_p(x,y)=\Vert x-H(x,y) \Vert_p$ (see Fig.~\ref{fig:pykeogh-2}),
then $\text{LB\_Keogh}_p(x,y)$ is a lower bound to $\text{NDTW}_p(x,y)$ and thus $\text{DTW}_p(x,y)$.
The following corollary follows  from Theorem~\ref{thm:mainthm} and Proposition~\ref{prop:error}.

\begin{corollary}
Given two equal-length time series $x$ and $y$ and $1\leq p\leq\infty$, then
\begin{itemize}
\item $\text{LB\_Keogh}_p(x,y)$  is a lower bound
to the DTW: \begin{align*}\text{DTW}_p(x,y)  \geq \text{NDTW}_p(x,y) \geq   \text{LB\_Keogh}_p(x,y);\end{align*}
\item the accuracy of LB\_Keogh is bounded by the distance to the envelope: \begin{align*}\text{DTW}_p(x,y)-\text{LB\_Keogh}_p(x,y) \leq \Vert \max \{U(y)_i - y_i, y_i-L(y)_i\}_i \Vert_p\end{align*} for all $x$. 
\end{itemize}
\end{corollary}

%

Algorithm~\ref{algo:lbkeogh} shows how LB\_Keogh can be used
to find a nearest neighbor in a time series database.
We used $\textrm{DTW}_1$ for all implementations (see
Appendix~\ref{sec:whichmeasure}).
The computation of the envelope of the query time series is done
once (see line~\ref{line:mylbkeoghenvelope}).
The lower bound is computed in lines~\ref{line:mylbkeoghstart}
to \ref{line:mylbkeoghend}. If the lower bound is sufficiently
large, the DTW is not computed (see line~\ref{line:mylbkeoghtest}). 
Ignoring the computation of the full DTW, at most
$(2N+3)n$~comparisons between data points are required to process a database
containing $N$~time series.

\begin{algorithm}
\begin{small}
 \begin{algorithmic}[1]
\STATE \textbf{input} a time series $y$ indexed from $1$ to $n$
\STATE \textbf{input} a set $S$ of candidate time series
\RETURN the nearest neighbor $B$ to $y$ in $S$ under $\text{DTW}_1$
\STATE $U,L \leftarrow \text{envelope}(y)$\label{line:mylbkeoghenvelope}
\STATE $b\leftarrow \infty$ \COMMENT{$b$ stores $\min_{x \in S} \text{DTW}_1(x,y)$}
\FOR {candidate $x$ in $S$}
\STATE $\beta \leftarrow 0$ \COMMENT{$\beta$ stores the lower bound}  \label{line:mylbkeoghstart}
\FOR{$i \in \{1,2,\ldots,n\}$}
\IF{$x_i > U_i$}
\STATE $\beta \leftarrow \beta + x_i-U_i$
\ELSIF{$x_i < L_i$}
\STATE $\beta \leftarrow \beta + L_i-x_i$\label{line:mylbkeoghend}
\ENDIF
\ENDFOR
\IF{$\beta < b$ }\label{line:mylbkeoghtest}
\STATE $t\leftarrow \text{DTW}_1(a,c)$  \COMMENT{We compute the full DTW.}
\IF{$t < b$}
\STATE $b\leftarrow t$
\STATE $B\leftarrow c$
\ENDIF
\ENDIF
\ENDFOR
 \end{algorithmic}
\end{small}
\caption{\label{algo:lbkeogh}LB\_Keogh-based Nearest-Neighbor algorithm
}
\end{algorithm}

\section{LB\_Improved}

\label{sec:lbimproved}
\begin{figure}
\centering
\includegraphics[width=0.7\columnwidth]{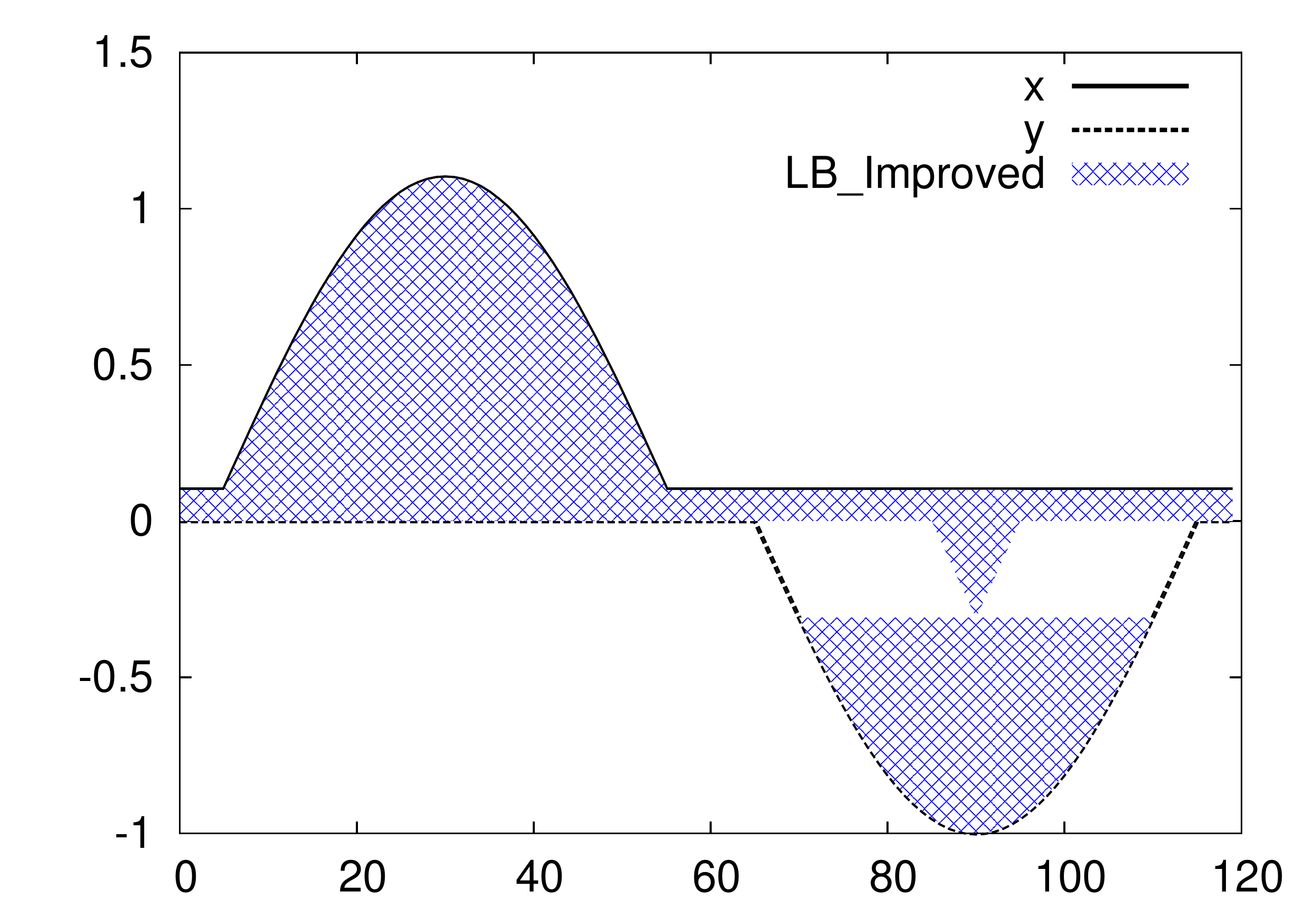}
\caption{\label{fig:pykeogh-3}LB\_Improved example: the area of the marked region is $\text{LB\_Improved}_1(x,y)$}
\end{figure}

In the previous Section, we saw that   $\text{NDTW}_p(x,y)^p\geq \text{LB\_Keogh}_p(x,y)^p + \textrm{NDTW}_p(H(x,y),y)^p$ for $1 \leq p<\infty$.
In turn, we have $\textrm{NDTW}_p(H(x,y),y) \geq \text{LB\_Keogh}_p(y,H(x,y))$.
Hence, write \[\text{LB\_Improved}_p(x,y)^p=\text{LB\_Keogh}_p(x,y)^p+\text{LB\_Keogh}_p(y,H(x,y))^p\]
for $1 \leq p<\infty$. 
By definition, we have  $\text{LB\_Improved}_p(x,y)\geq \text{LB\_Keogh}_p(x,y)$.
Intuitively, whereas $\text{LB\_Keogh}_p(x,y)$ measures the distance between $x$ and the envelope of $y$,
$\text{LB\_Keogh}_p(y,H(x,y))$ measures the distance between $y$ and the envelope of the projection
of $x$ on $y$  (see Fig.~\ref{fig:pykeogh-3}).
The next corollary shows that LB\_Improved is a lower bound to the DTW\@.

\begin{corollary}
Given two equal-length time series $x$ and $y$ and $1\leq p<\infty$, then
 $\text{LB\_Improved}_p(x,y)$  is a lower bound
to the DTW: $\text{DTW}_p(x,y)  \geq \text{NDTW}_p(x,y) \geq   \text{LB\_Improved}_p(x,y)$.
\end{corollary}
\begin{pf}
%
Recall that $\text{LB\_Keogh}_p(x,y)=\Vert x-H(x,y) \Vert_p$. 
First apply Theorem~\ref{thm:mainthm}: $ \text{DTW}_p(x,y)^p  \geq \text{NDTW}_p(x,y)^p  \geq   \text{LB\_Keogh}_p(x,y)^p + \text{NDTW}_p(H(x,y),y)^p$.
Apply Theorem~\ref{thm:mainthm} once more:
$ \text{NDTW}_p(y,H(x,y))^p  \geq   \text{LB\_Keogh}_p(y,H(x,y))^p$. 
By substitution, we get
$ \text{DTW}_p(x,y)^p  \geq \text{NDTW}_p(x,y)^p  \geq   \text{LB\_Keogh}_p(x,y)^p + \text{LB\_Keogh}_p(y,H(x,y))^p $ thus proving the result.
\end{pf}

\begin{figure*}  \centering
\subfloat[We begin with  $y$ and its envelope $L(y),U(y)$.]{\includegraphics[width=0.45\textwidth]{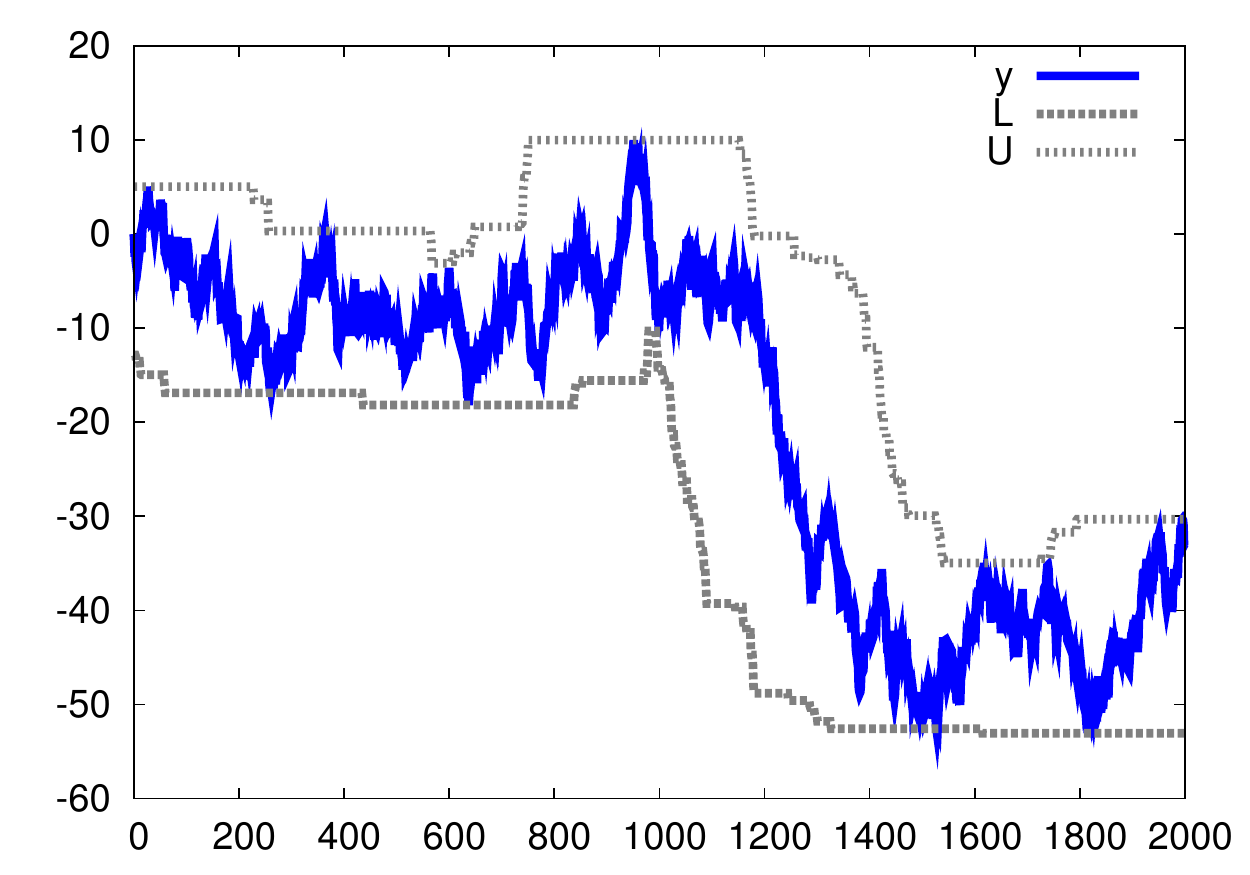}}\hspace*{0.5cm}
 \subfloat[We compare candidate $x$ with the envelope $L(y),U(y)$.]{\includegraphics[width=0.45\textwidth]{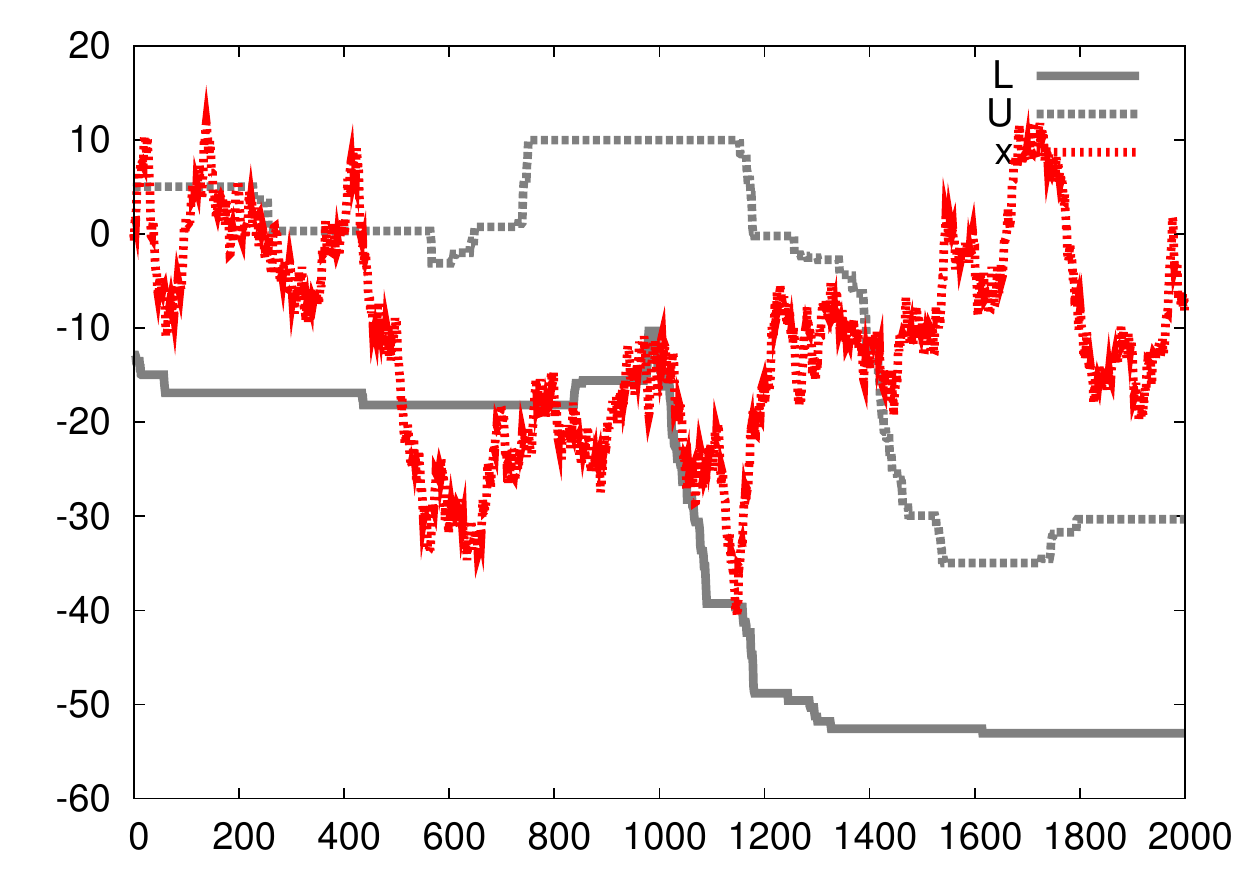}}\\ 
 \subfloat[The difference is LB\_Keogh($x$,$y$).]{\includegraphics[width=0.45\textwidth]{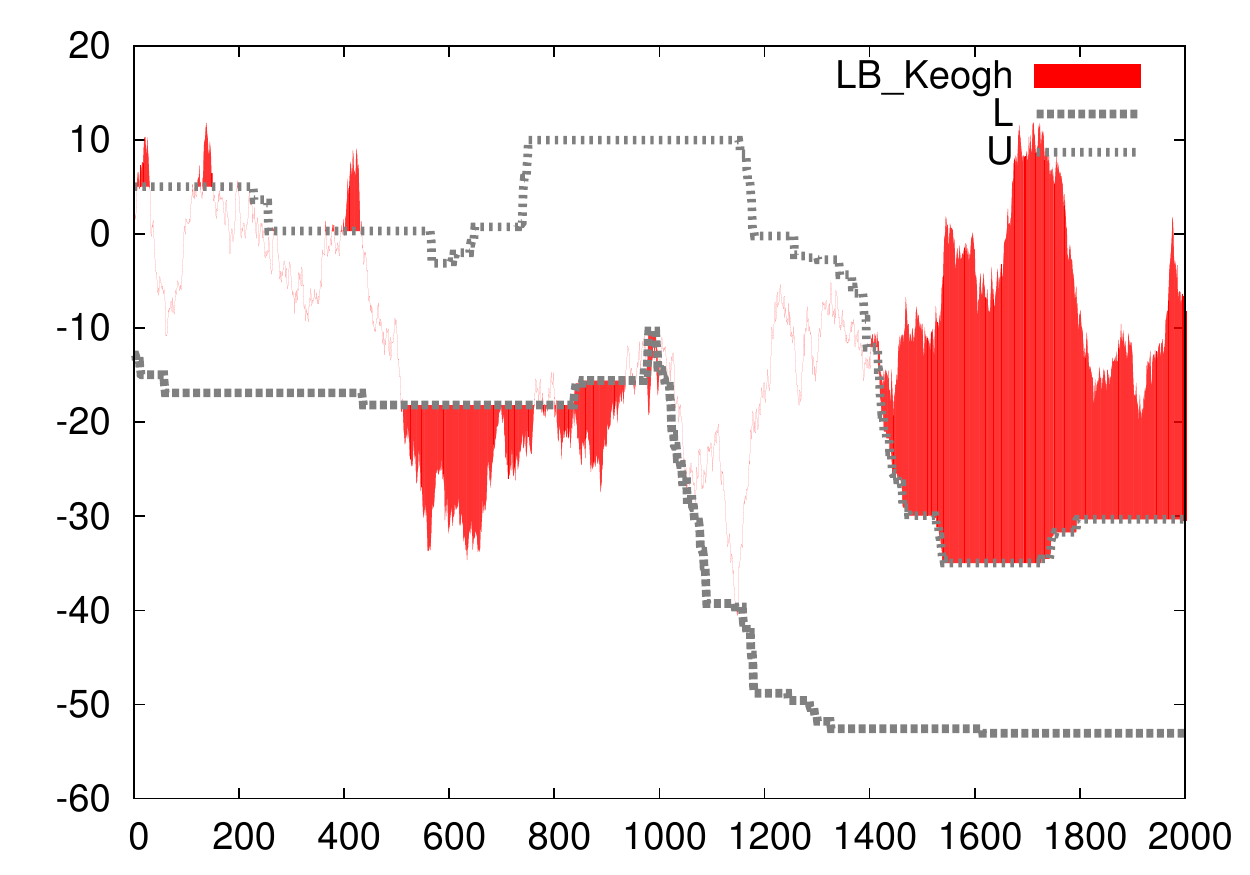}}
 \hspace*{0.5cm}\subfloat[We compute $x'$, the projection of $x$ on the envelope $L(y),U(y)$.]{\includegraphics[width=0.45\textwidth]{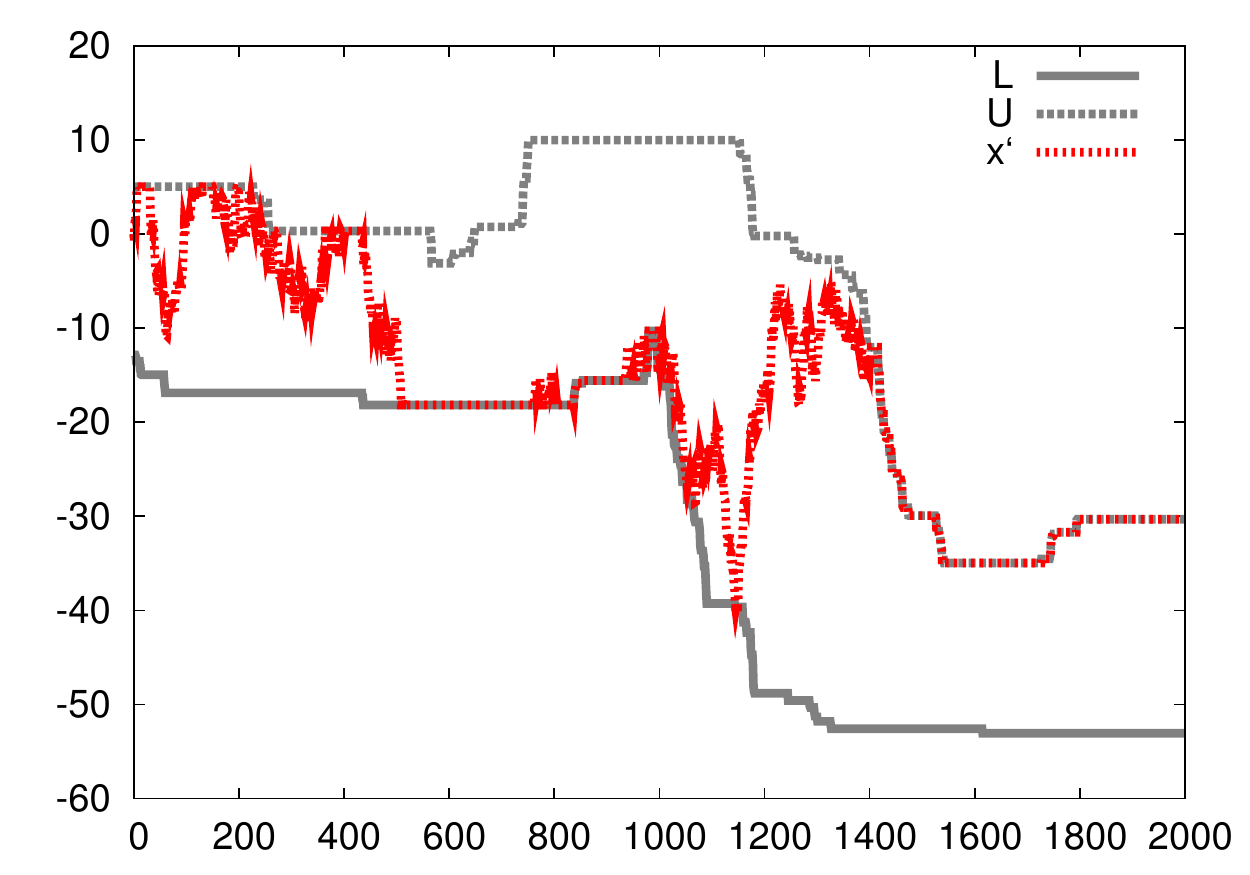}} \\
\subfloat[We compute the envelope of $x'$.]{\includegraphics[width=0.45\textwidth]{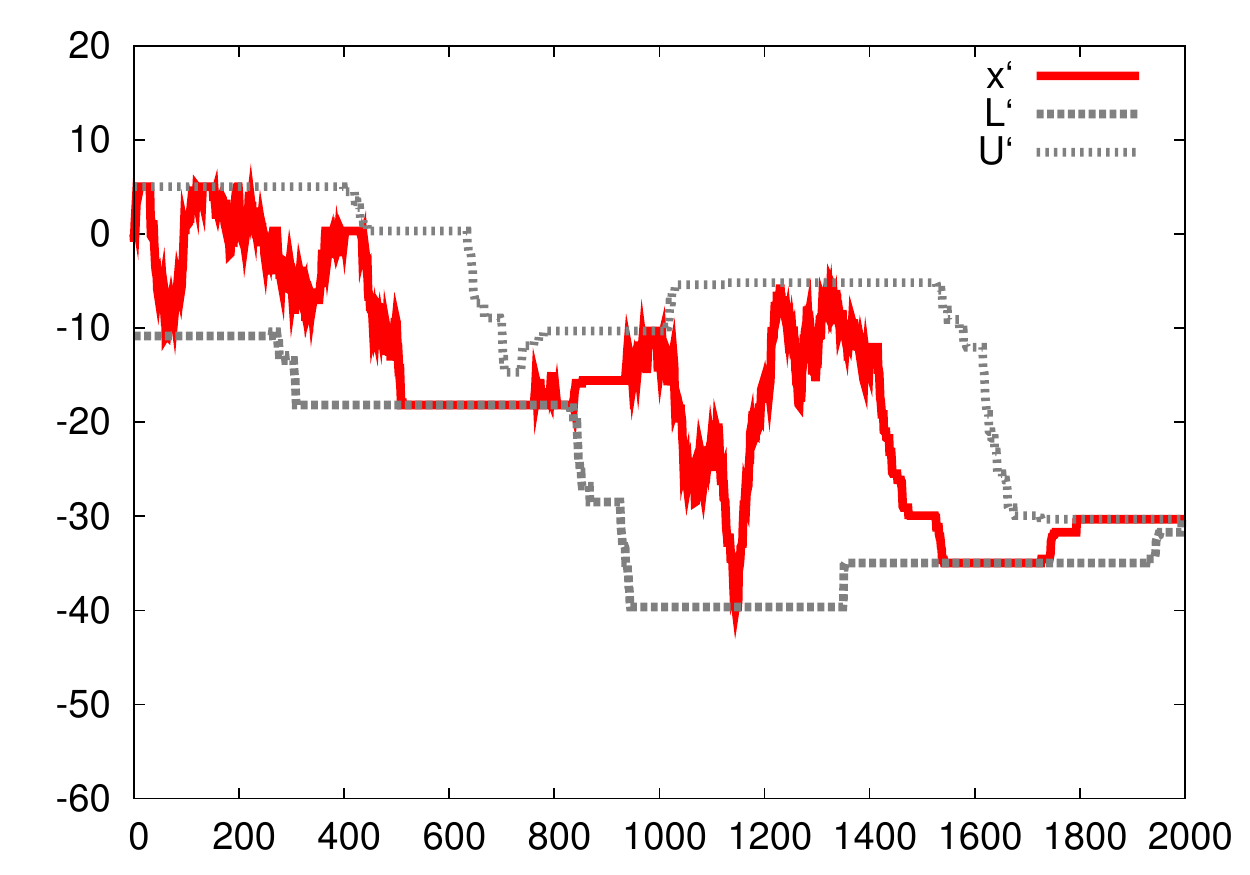}}
\hspace*{0.5cm} \subfloat[The difference between $y$ and the envelope $L(x'),U(x')$
 is added to LB\_Keogh to compute LB\_Improved.]{\includegraphics[width=0.45\textwidth]{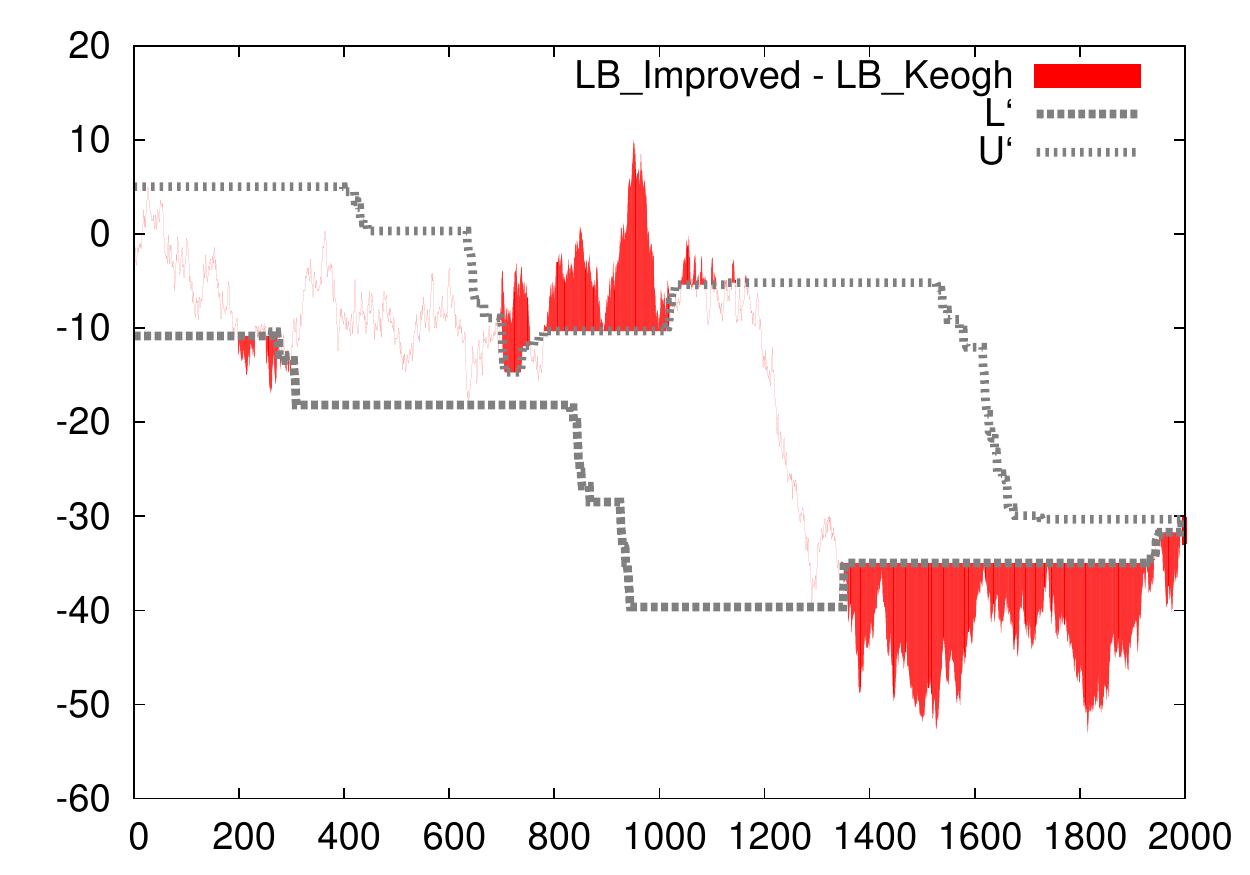}} 
\caption{\label{fancyplot}Computation of LB\_Improved as in Algorithm~\ref{algo:lbimproved}}
\end{figure*}

Algorithm~\ref{algo:lbimproved} shows how to apply LB\_Improved as a two-step process (see Fig.~\ref{fancyplot}). Initially, for each candidate $x$,
we compute the lower bound $\text{LB\_Keogh}_{1}(x,y)$ (see lines~\ref{line:lbkeoghstart} to \ref{line:lbkeoghend}). If this lower bound  is sufficiently large, the candidate is 
discarded (see line~\ref{line:lbkeoghdiscard}), otherwise we add $\text{LB\_Keogh}_1(y,H(x,y))$ to $\text{LB\_Keogh}_{1}(x,y)$, in effect computing $\text{LB\_Improved}_1(x,y)$ (see lines~\ref{line:lbimprovedstart} to \ref{line:lbimprovedend}). If this larger lower bound is sufficiently large, the candidate
is finally discarded (see line~\ref{line:lbimproveddiscard}). Otherwise, we compute the full DTW\@.
If $\alpha$ is the fraction of candidates pruned by LB\_Keogh, at most
$(2N+3)n+5(1-\alpha)Nn$~comparisons between data points are required to process a database
containing $N$~time series.

\begin{algorithm}
\begin{small}
 \begin{algorithmic}[1]
\STATE \textbf{input} a time series $y$ indexed from $1$ to $n$
\STATE \textbf{input} a set $S$ of candidate time series
\RETURN the nearest neighbor $B$ to $y$ in $S$ under $\text{DTW}_1$
\STATE $U,L \leftarrow \text{envelope}(y)$
\STATE $b\leftarrow \infty$ \COMMENT{$b$ stores $\min_{x \in S} \text{DTW}_1(x,y)$}
\FOR {candidate $x$ in $S$}
\STATE copy $x$ to $x'$\label{line:copy} \COMMENT{$x'$ will store the projection of $x$ on $y$} 
\STATE $\beta \leftarrow 0$  \COMMENT{$\beta$ stores the lower bound}  \label{line:lbkeoghstart}
\FOR{$i \in \{1,2,\ldots,n\}$}
\IF{$x_i > U_i$}
\STATE $\beta \leftarrow \beta + x_i-U_i$
\STATE $x'_i = U_i$\label{line:modif1}
\ELSIF{$x_i < L_i$}
\STATE $\beta \leftarrow \beta + L_i-x_i$
\STATE $x'_i = L_i$\label{line:lbkeoghend}\label{line:modif2}
\ENDIF
\ENDFOR
\IF{$\beta < b$}\label{line:lbkeoghdiscard}
\STATE $U',L' \leftarrow \text{envelope}(x')$  \label{line:lbimprovedstart}
\FOR{$i \in \{1,2,\ldots,n\}$}
\IF{$y_i > U'_i$}
\STATE $\beta\leftarrow \beta + y_i- U'_i$
\ELSIF{$y_i < L'_i$}
\STATE $\beta\leftarrow \beta + L'_i- y_i$\label{line:lbimprovedend}
\ENDIF
\ENDFOR
\IF{$\beta < b$}\label{line:lbimproveddiscard}
\STATE $t\leftarrow \text{DTW}_1(a,c)$  \COMMENT{We compute the full DTW.}
\IF{$t < b$}
\STATE $b\leftarrow t$
\STATE $B\leftarrow c$
\ENDIF
\ENDIF
\ENDIF
\ENDFOR
 \end{algorithmic}
\end{small}
\caption{\label{algo:lbimproved}LB\_Improved-based Nearest-Neighbor algorithm
}
\end{algorithm}

\section{Using a multidimensional indexing structure}
\label{sec:multiindex}
The running time of Algorithms~\ref{algo:lbkeogh} and \ref{algo:lbimproved} may be improved if 
we use a multidimensional index such as an R*-tree~\cite{beckmann1990rte}.
Unfortunately, the performance of such an index  diminishes quickly as the number
of dimensions increases~\cite{671192}. 
To solve this problem, 
several dimensionality reduction techniques are possible such
as piecewise 
linear~\cite{xiao2004nst,shou2005,dong2006lsd} segmentation.
Following Zhu and Shasha~\cite{zhu2003wie}, we 
project time series and their envelopes on a $d$-dimensional space
using piecewise sums:
$P_d(x)= (\sum_{i\in C_j} x_i)_j$ where $C_1, C_2,\ldots, C_d$ is a disjoint
cover of $\{1,2,\ldots, n\}$. Unlike Zhu and Shasha, we do
not require the intervals to have equal length. 
The $l_1$ distance between $P_d(y)$ and the minimum bounding hyperrectangle containing
$P_d(L(x))$ and $P_d(U(x))$
is a lower bound to the $\textrm{DTW}_1(x,y)$: 
\begin{align*}
 \textrm{DTW}_1(x,y) & \geq  \textrm{LB\_Keogh}_1(x,y) \\
 & =  \sum_{i=1}^n d(x_i, [L(y)_i, U(y)_i]) \\
& \geq   \sum_{j=1}^d d(  P_d(x)_j,[P_d(L(y))_j, P_d(U(y))_j] ).
\end{align*}
For our experiments, 
we chose
the cover $C_j=[1+(j-1) \lfloor n/d \rfloor,
j \lfloor n/d \rfloor]$ for $j=1,\ldots, d-1$ and
$C_d=  [1+(d-1) \lfloor n/d \rfloor,n]$.

We can summarize the Zhu-Shasha R*-tree algorithm as
follows:
\begin{enumerate}
\item for each time series $x$ in the database,
add $P_d(x)$ to the R*-tree;
\item given a query time series $y$, compute
its envelope $E=P_d(L(y)),P_d(U(y))$;
\item starting with $b = \infty$,
iterate over all candidate $P_d(x)$ at a $l_1$ distance $b$ from the envelope
$E$ using the R*-tree, once a candidate
is found, update $b$ with $\textrm{DTW}_1(x,y)$ and repeat
until you have exhausted all candidates.
\end{enumerate}
This algorithm is correct because the distance between
$E$ and $P_d(x)$ is a lower bound to $\textrm{DTW}_1(x,y)$.  
However, dimensionality reduction diminishes the pruning
power of LB\_Keogh : $d(E,P_d(x)) \leq  \textrm{LB\_Keogh}_1(x,y)$.
Hence, we propose a new algorithm (\textsc{R*-Tree+LB\_Keogh}) where instead
of immediately updating 
$b$ with $\textrm{DTW}_1(x,y)$, we first compute
the LB\_Keogh 
 lower bound between $x$ and $y$. 
Only when it is less than $b$, do we compute the
full DTW. Finally, as a third algorithm
(\textsc{R*-Tree+LB\_Improved}), we first compute LB\_Keogh,
and if it is less than $b$, then we compute
LB\_Improved, and only when it is also lower than $b$
do we compute the DTW, as in Algorithm~\ref{algo:lbimproved}. 
\textsc{R*-tree}+\textsc{LB\_Improved} has maximal  pruning power,
whereas Zhu-Shasha R*-tree has the lesser pruning power of the three alternatives.

\section{Comparing Zhu-Shasha R*-tree, LB\_Keogh, and LB\_Improved}
\label{sec:comparing}
%
%
%
%
 %

In this section, we benchmark algorithms Zhu-Shasha R*-tree, \textsc{R*-tree}+ \textsc{LB\_Keogh}, and \textsc{R*-tree}+\textsc{LB\_Improved}. 
We know that the LB\_Improved approach  has at least the pruning power of the other methods,
but does more pruning translate into a faster nearest-neighbor retrieval under the  DTW distance?

We implemented the algorithms in C++ using an external-memory R*-tree. The time series are stored on disk in a binary flat file.
We used the GNU GCC~4.0.2 compiler on  an Apple Mac~Pro, 
having two  Intel Xeon dual-core processors running at 2.66\,GHz with 2\,GiB of RAM\@. 
No thrashing was observed.
We measured  the wall-clock total time.
In all experiments, we benchmark
nearest-neighbor retrieval under the $\text{DTW}_1$.
By default, the locality constraint $w$ is set at 10\% ($w=n/10$).
To ensure reproducibility, our source code is freely available~\cite{googlelbimproved},
including the script used to generate synthetic data sets. We compute the
full DTW using a $O(nw)$-time dynamic programming algorithm.

The R*-tree was implemented using the Spatial Index library~\cite{spatialindex}. In informal tests,
we found that
a projection on an 8-dimensional space, as
described by Zhu and Shasha, gave good  results: substantially larger ($d>10$) or smaller ($d<6$)
settings gave poorer performance.
We used a 4,096-byte page size and a 10-entry internal
memory buffer. 

For \textsc{R*-tree}+ \textsc{LB\_Keogh} and \textsc{R*-tree}+\textsc{LB\_Improved},
we experimented with early abandoning~\cite{1106385} to cancel the computation of the
lower bound as soon as the error is too large. While it often improved retrieval time
slightly for both LB\_Keogh and LB\_Improved, the difference was always small (less than $\approx 1$\%).
One explanation is that the candidates produced by the Zhu-Shasha R*-tree are rarely
poor enough to warrant efficient early abandoning.

We do not report our benchmarking results over the simple 
Algorithms~\ref{algo:lbkeogh} and~\ref{algo:lbimproved}.
In almost all cases, the R*-tree equivalent---\textsc{R*-tree}+ \textsc{LB\_Keogh} or \textsc{R*-tree}+\textsc{LB\_Improved}---was at least slightly better  and sometimes several times faster.

\subsection{Synthetic data sets}
 
We tested our algorithms using the  Cylinder-Bell-Funnel~\cite{921732} and
 Control Charts~\cite{pham1998ccp} data sets, as well as over two databases of random walks.
 We generated 256-sample and 1~000-sample random-walk time series using the formula   $x_i = x_{i-1}+N(0,1)$ and $x_1=0$.
 
\begin{figure*}
\centering
  \subfloat[Average Retrieval Time]{\includegraphics[width=0.45\textwidth]{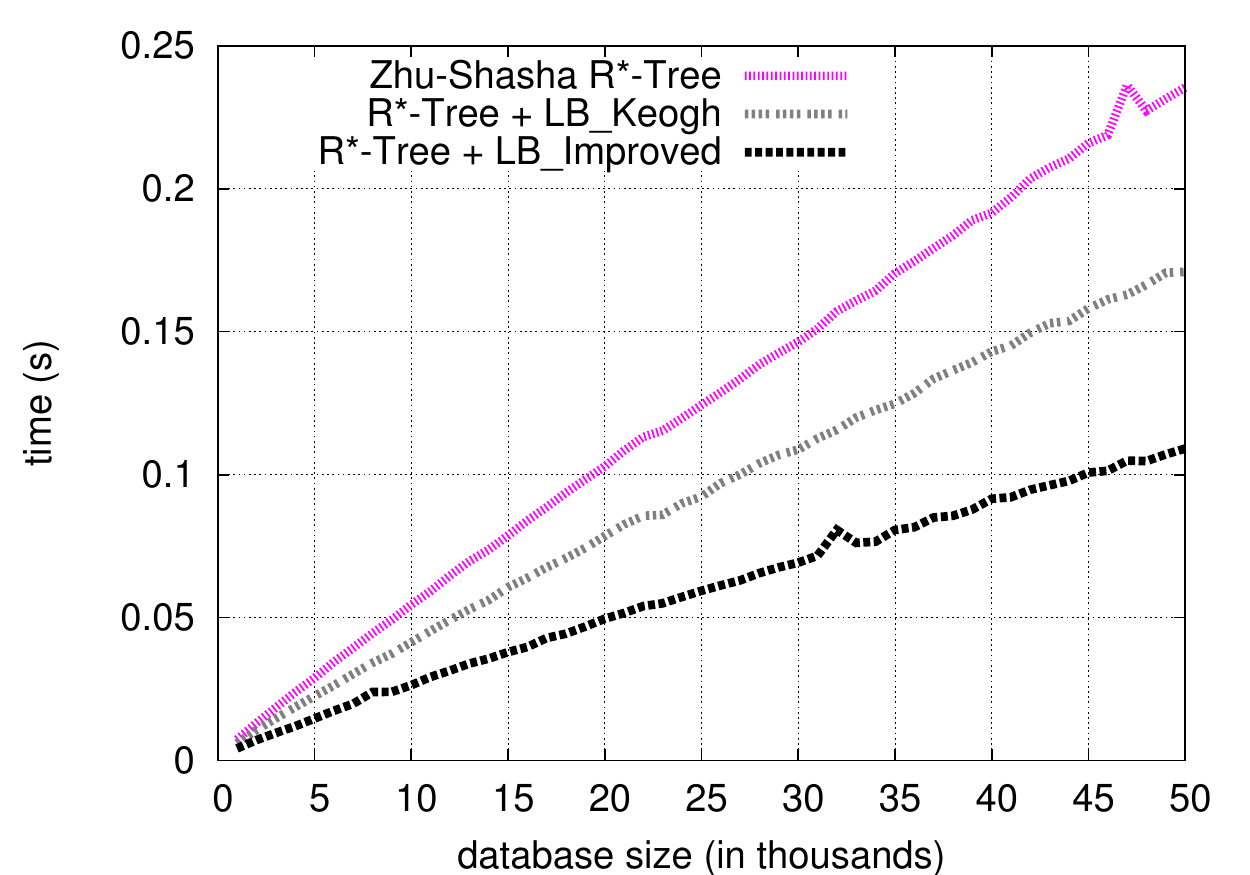}}
  \subfloat[Pruning Power]{\includegraphics[width=0.45\textwidth]{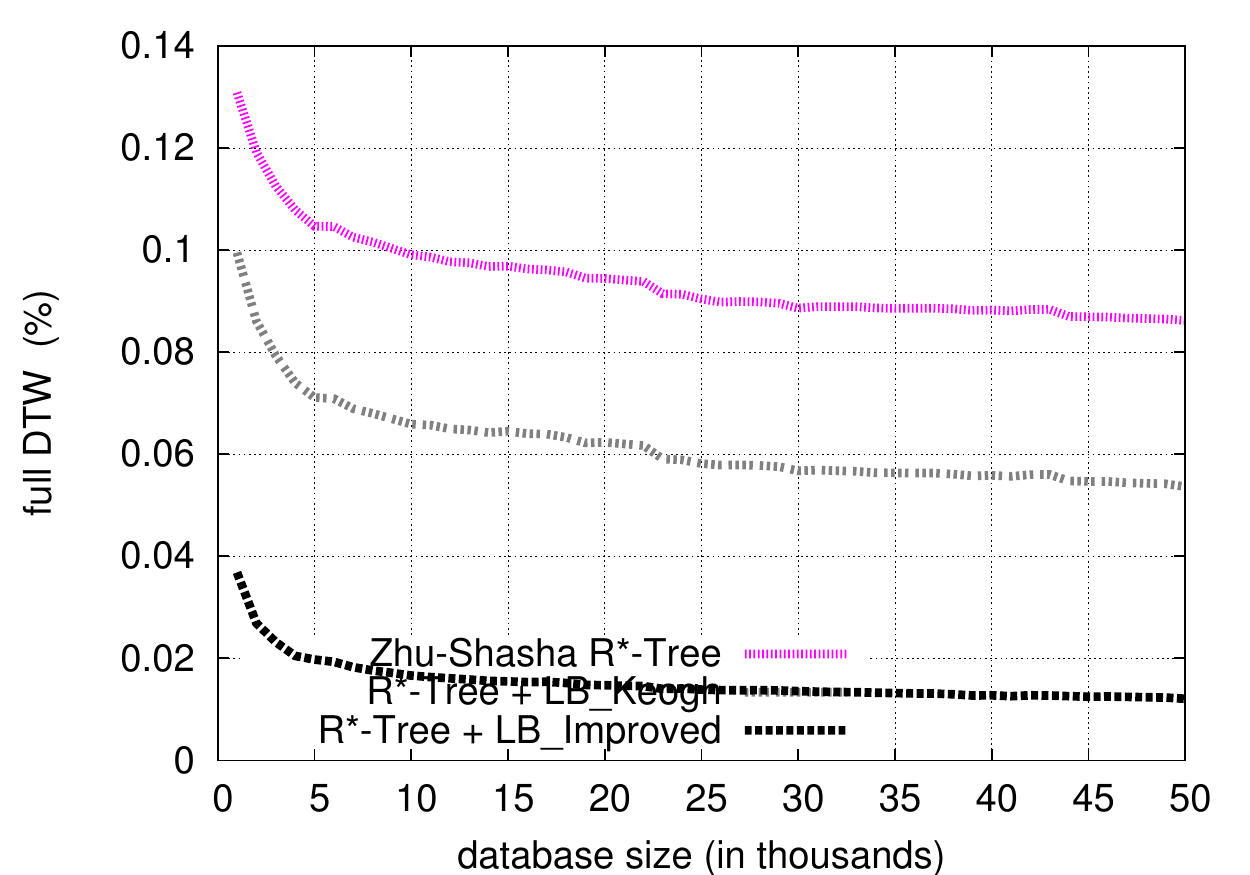}} 
\caption{Nearest-Neighbor Retrieval for the 256-sample random-walk data set\label{fig:perf0}}
\end{figure*}

\begin{figure*}
\centering
  \subfloat[Average Retrieval Time]{\includegraphics[width=0.45\textwidth]{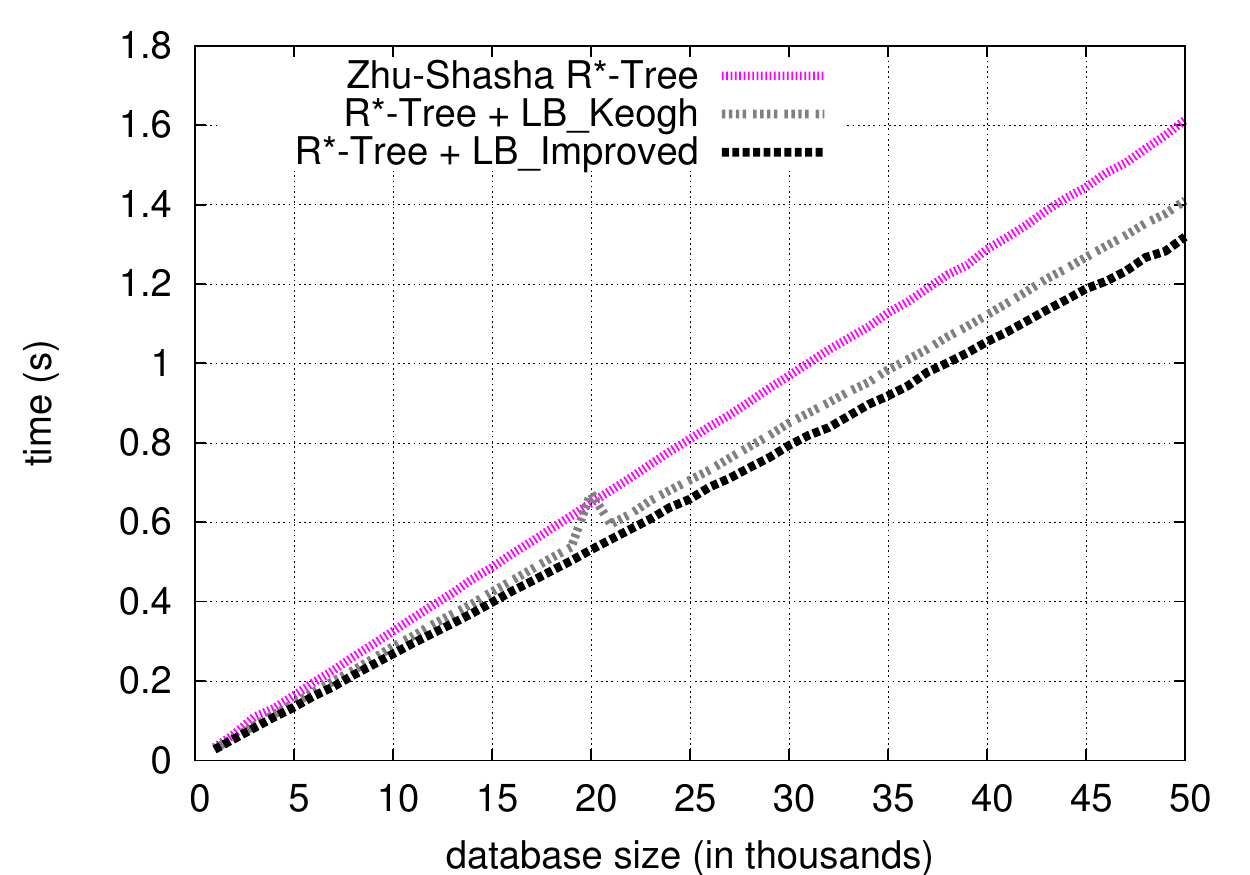}}
  \subfloat[Pruning Power]{\includegraphics[width=0.45\textwidth]{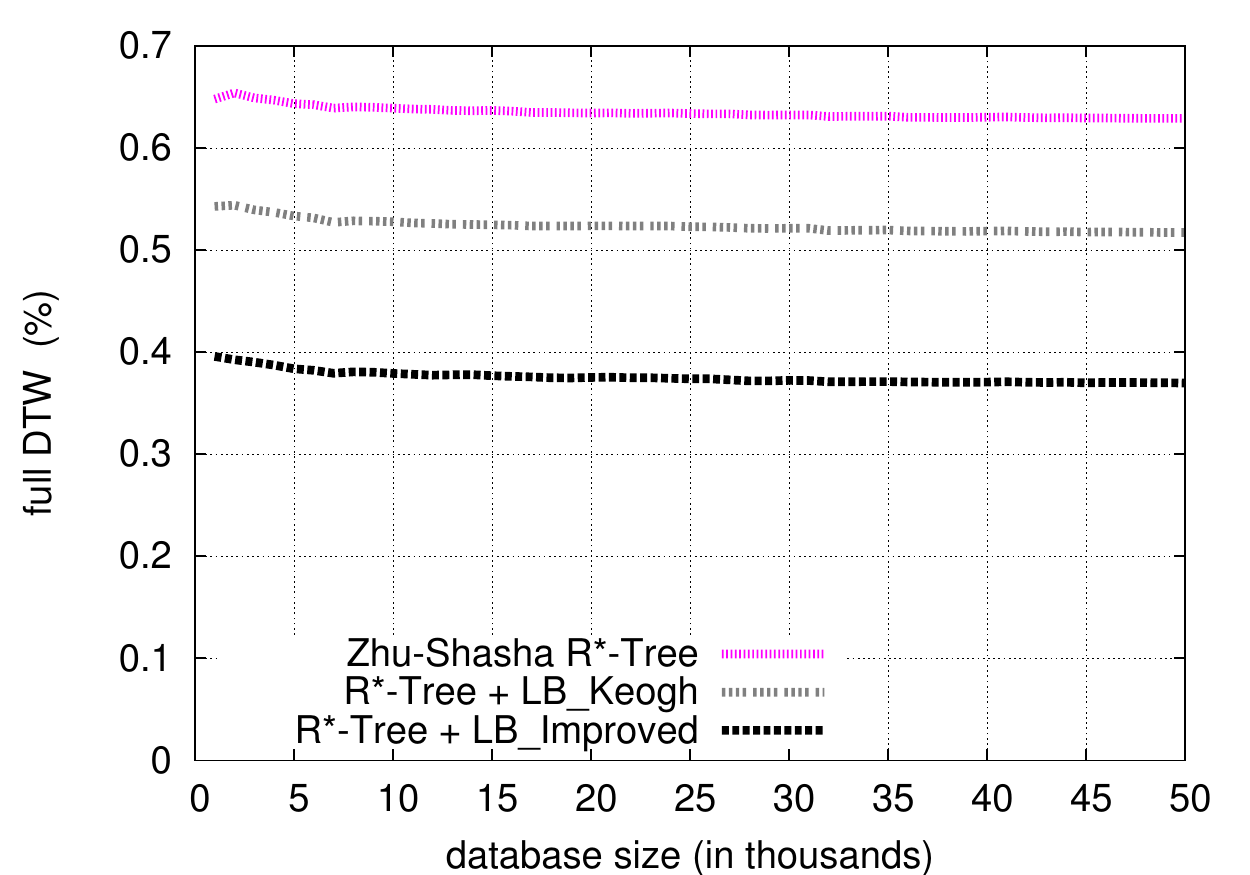}} 
\caption{Nearest-Neighbor Retrieval for the Cylinder-Bell-Funnel data set\label{fig:perf1}}
\end{figure*}

\begin{figure*}
\centering
  \subfloat[Average Retrieval Time]{\includegraphics[width=0.45\textwidth]{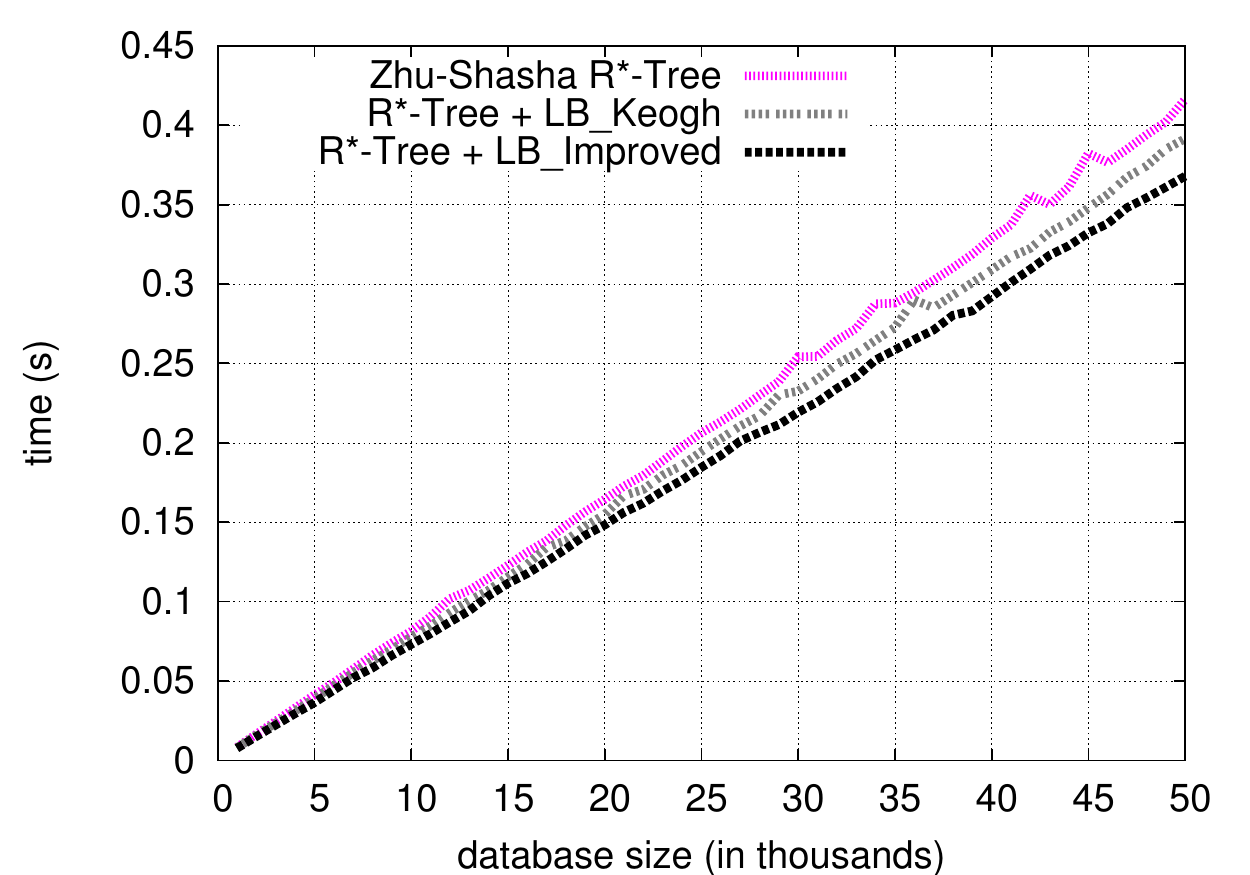}}
  \subfloat[Pruning Power]{\includegraphics[width=0.45\textwidth]{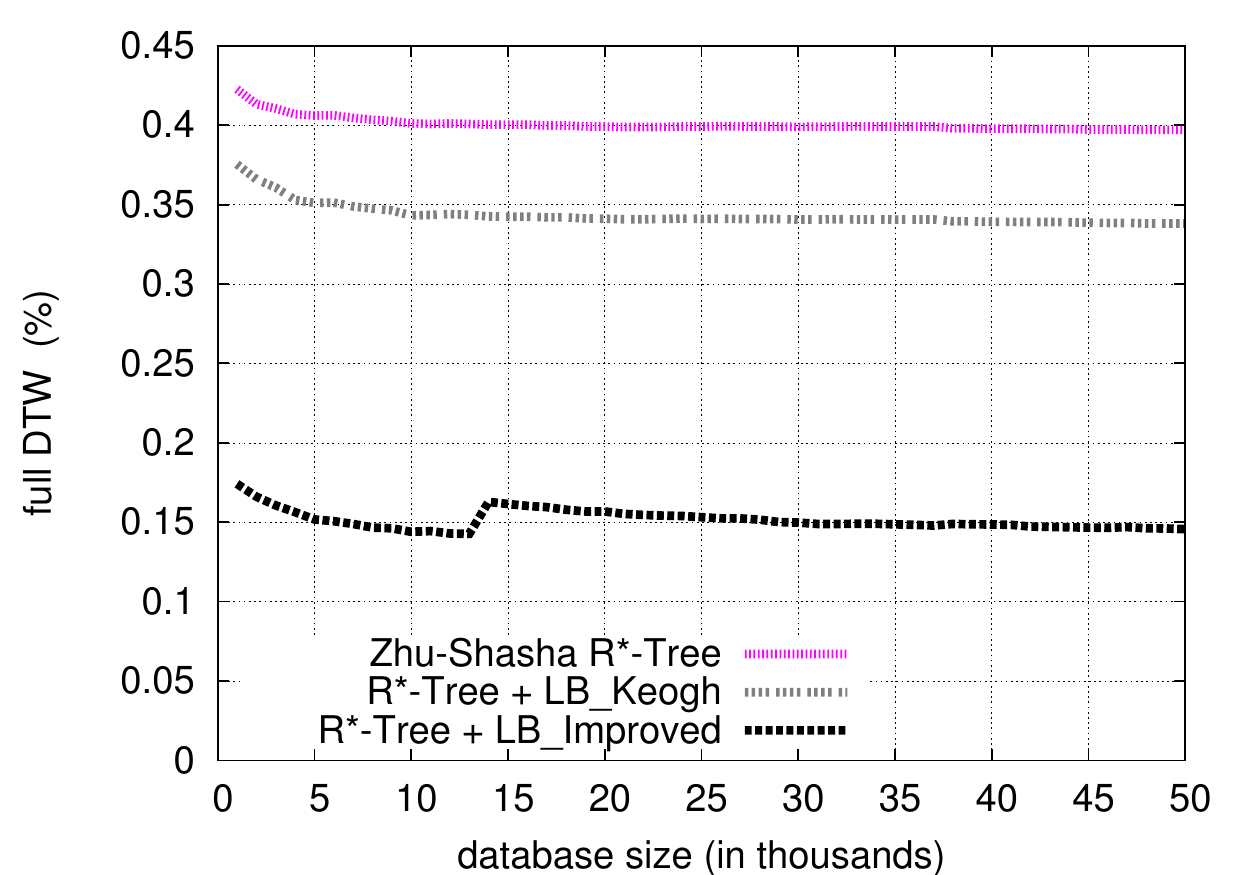}} 
\caption{Nearest-Neighbor Retrieval for the Control Charts data set\label{fig:perf2}}
\end{figure*}

\begin{figure*}
\centering
  \subfloat[Average Retrieval Time]{\includegraphics[width=0.45\textwidth]{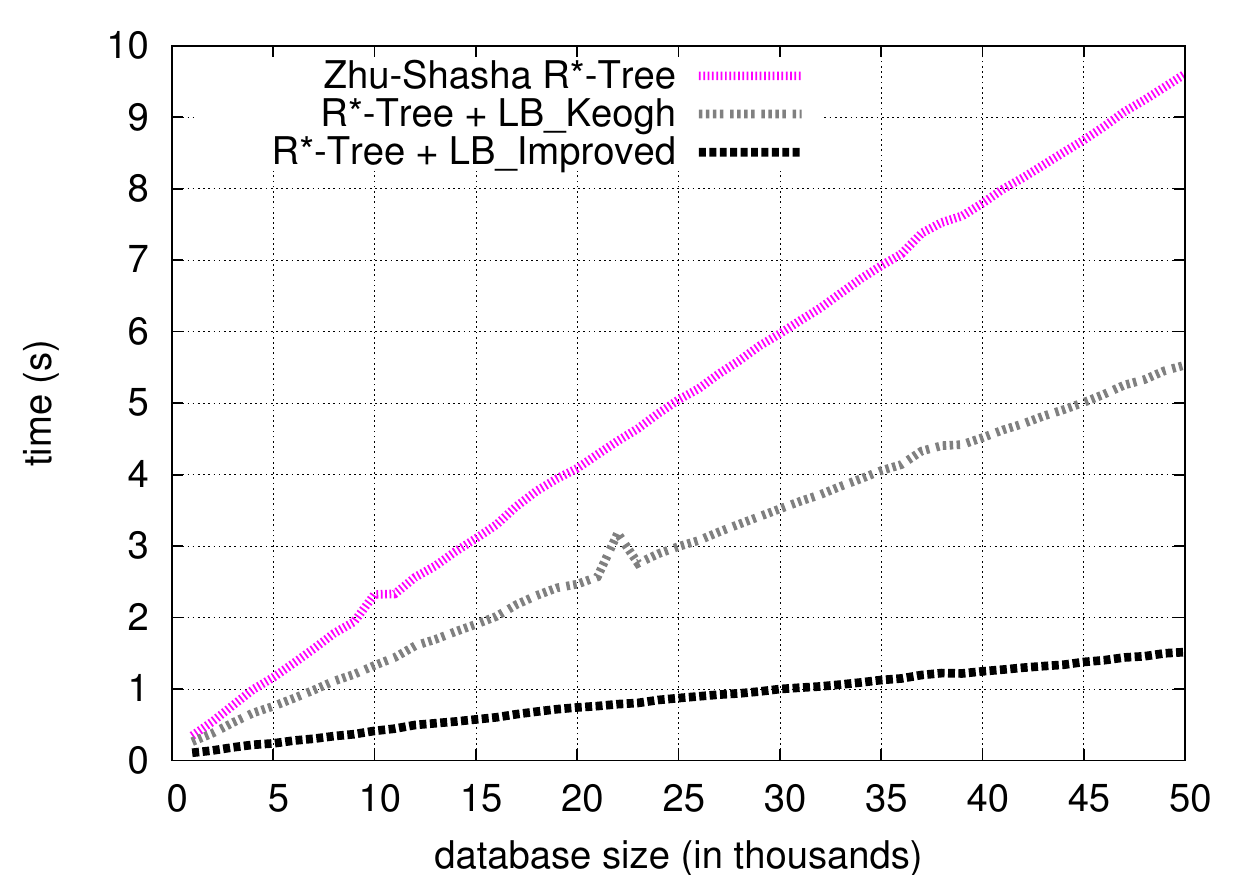}}
  \subfloat[Pruning Power]{\includegraphics[width=0.45\textwidth]{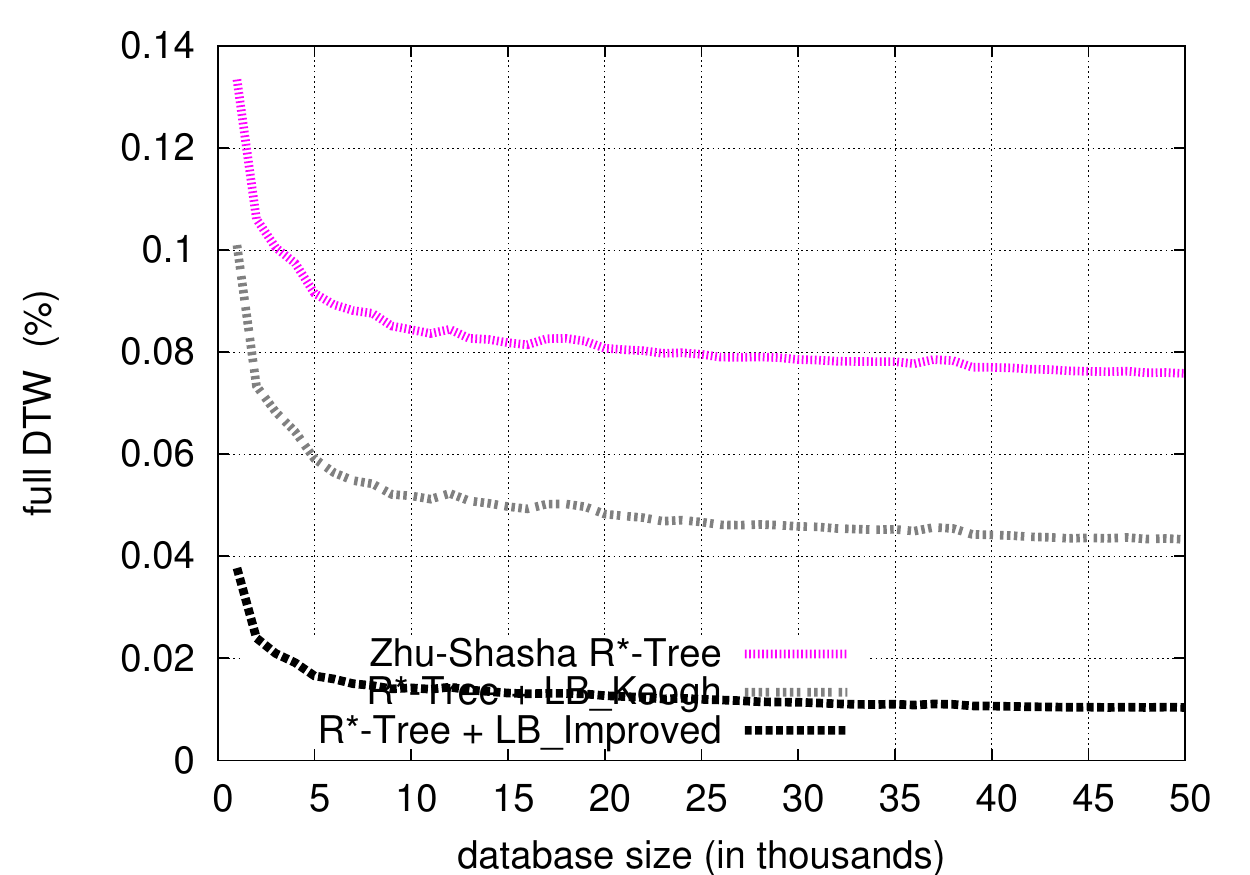}} 
\caption{Nearest-Neighbor Retrieval for the 1000-sample random-walk data set\label{fig:perf3}}
\end{figure*}

 For each data set, we generated a database of 50~000~time series by adding 
 randomly chosen items. 
 Figs.~\ref{fig:perf0}, \ref{fig:perf1}, \ref{fig:perf2} and \ref{fig:perf3} show
 the average timings and pruning ratio averaged over 20~queries based on 
 randomly chosen
 time series as we consider larger and large fraction of the database.
LB\_Improved  prunes between 2 and 4 times more candidates than 
 LB\_Keogh.
  \textsc{R*-tree+LB\_Improved}  
   is faster than Zhu-Shasha R*-tree by a factor between 0 and 6. 
 
 We saw almost no performance gain over Zhu-Shasha R*-tree with simple time series such as 
  the Cylinder-Bell-Funnel or the Control Charts data sets.
 However, in these cases, even LB\_Improved has  modest pruning powers of
  40\% and 15\%. Low pruning means that the computational cost is dominated
  by the cost of the full DTW.


\subsection{Shape data sets}

\begin{figure*}
\centering
  \subfloat[Average Retrieval Time]{\includegraphics[width=0.45\textwidth]{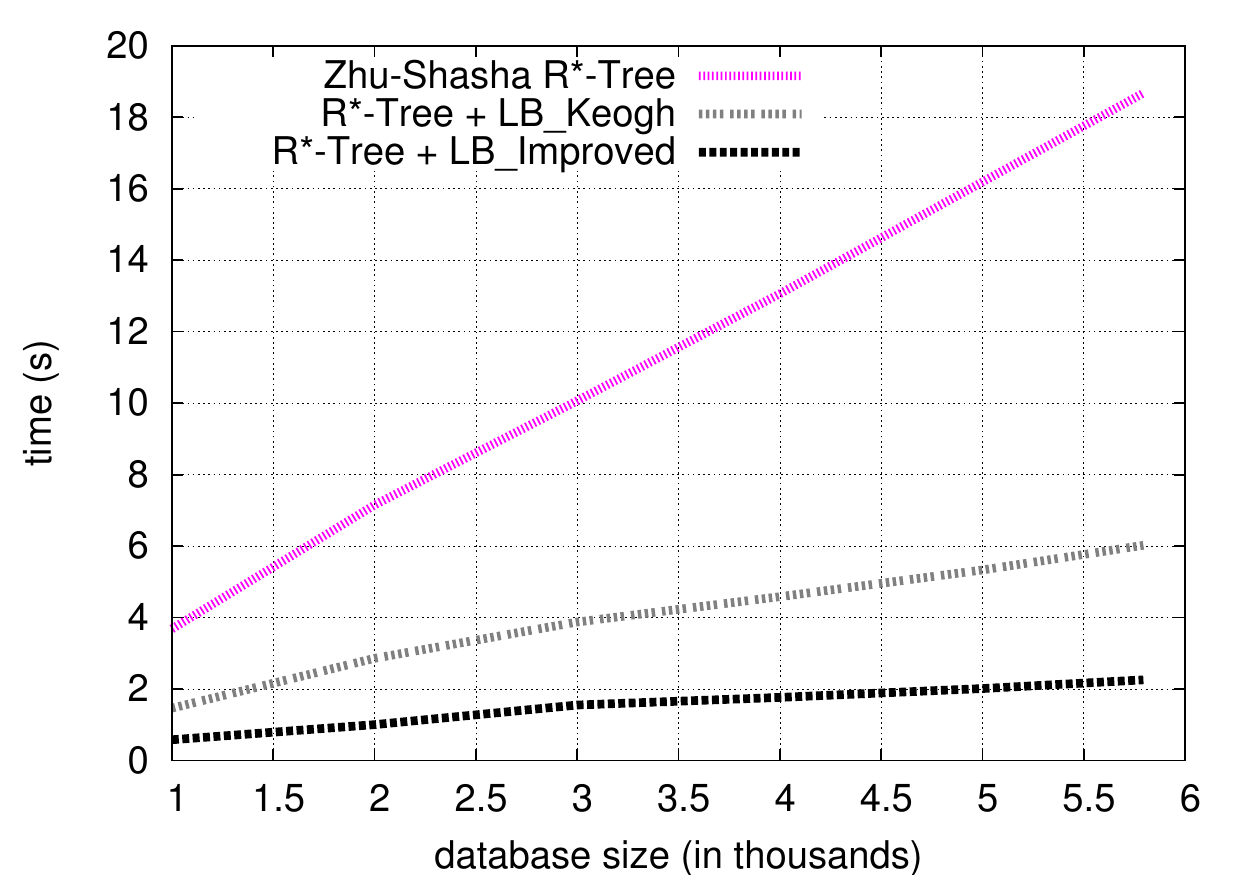}}
  \subfloat[Pruning Power]{\includegraphics[width=0.45\textwidth]{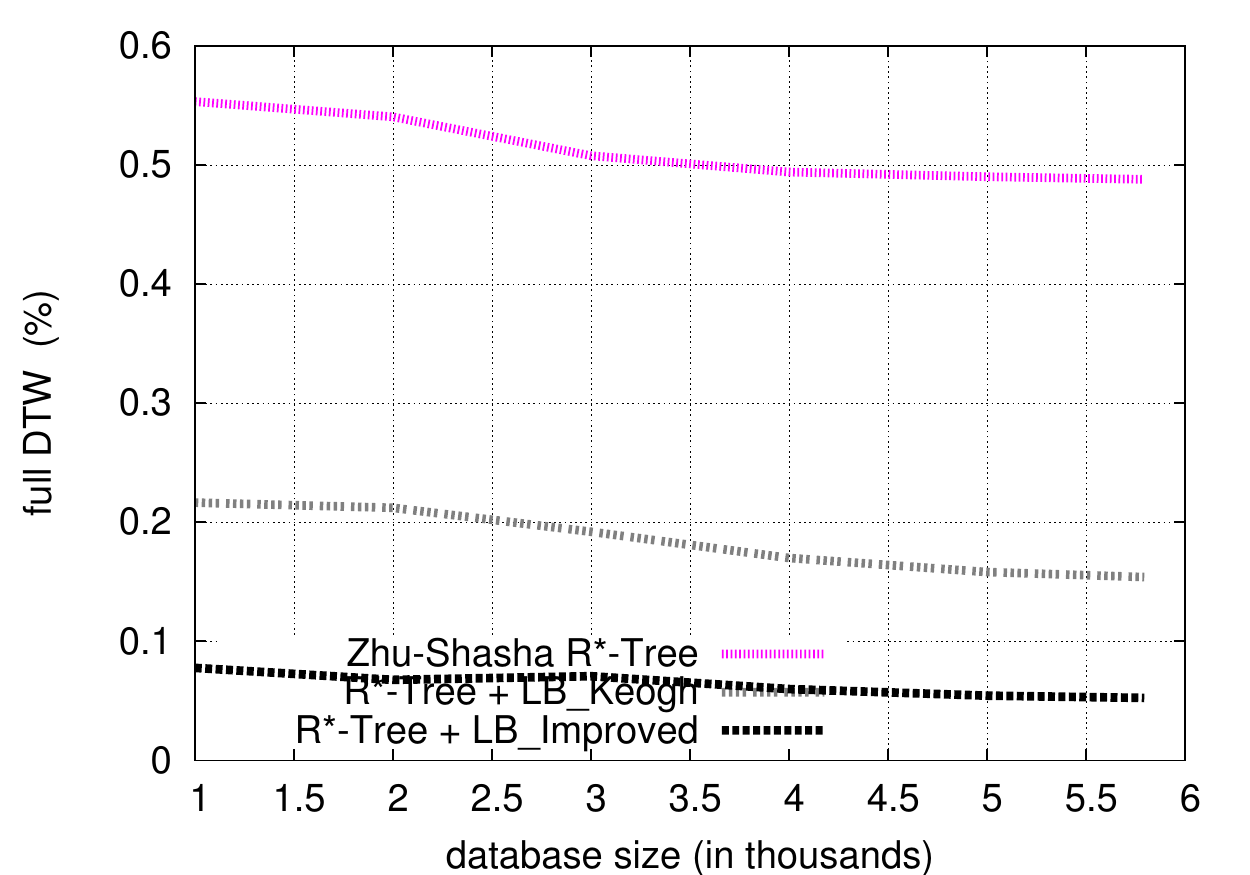}}
\caption{Nearest-Neighbor Retrieval for the heterogeneous shape data set\label{fig:perf4}}
\end{figure*}

\begin{figure*}
\centering
  \subfloat[Average Retrieval Time]{\includegraphics[width=0.45\textwidth]{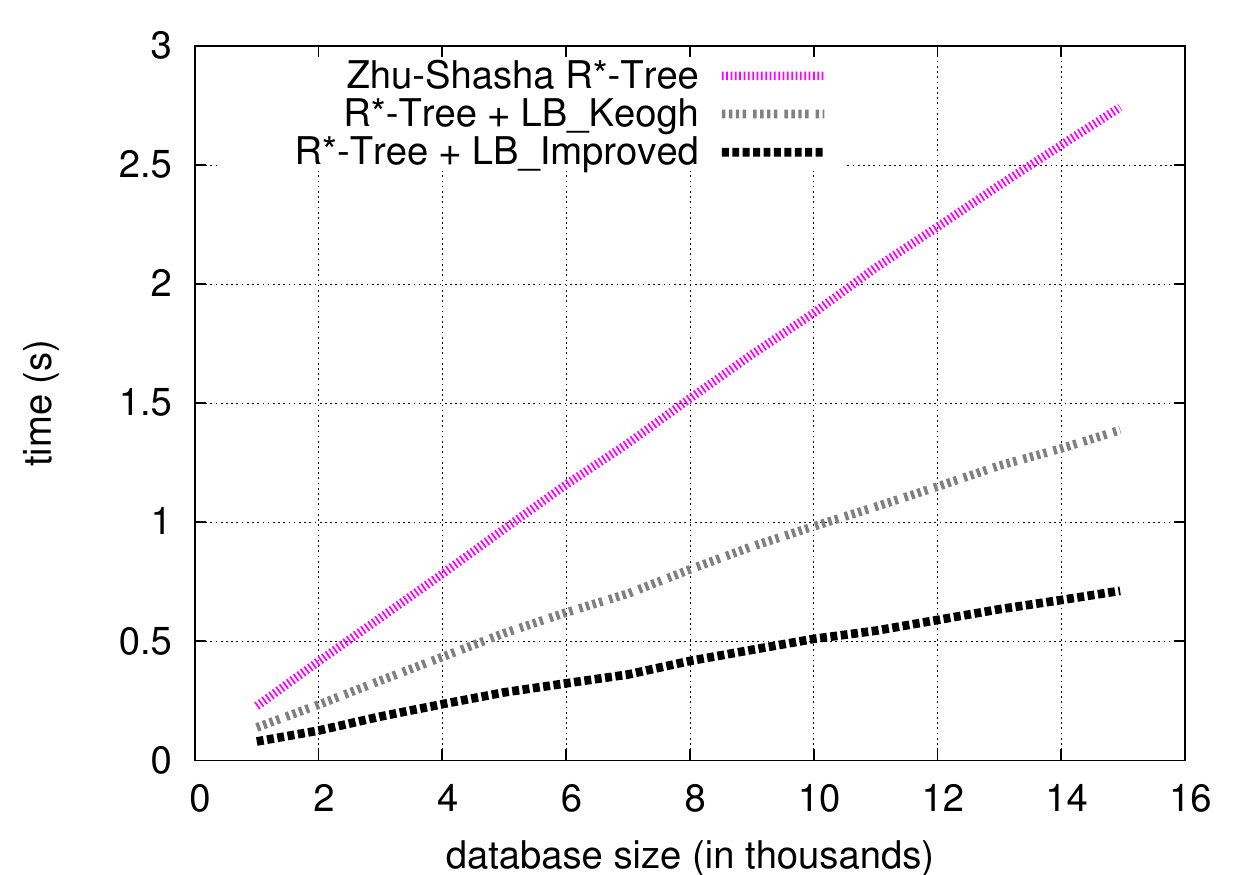}}
  \subfloat[Pruning Power]{\includegraphics[width=0.45\textwidth]{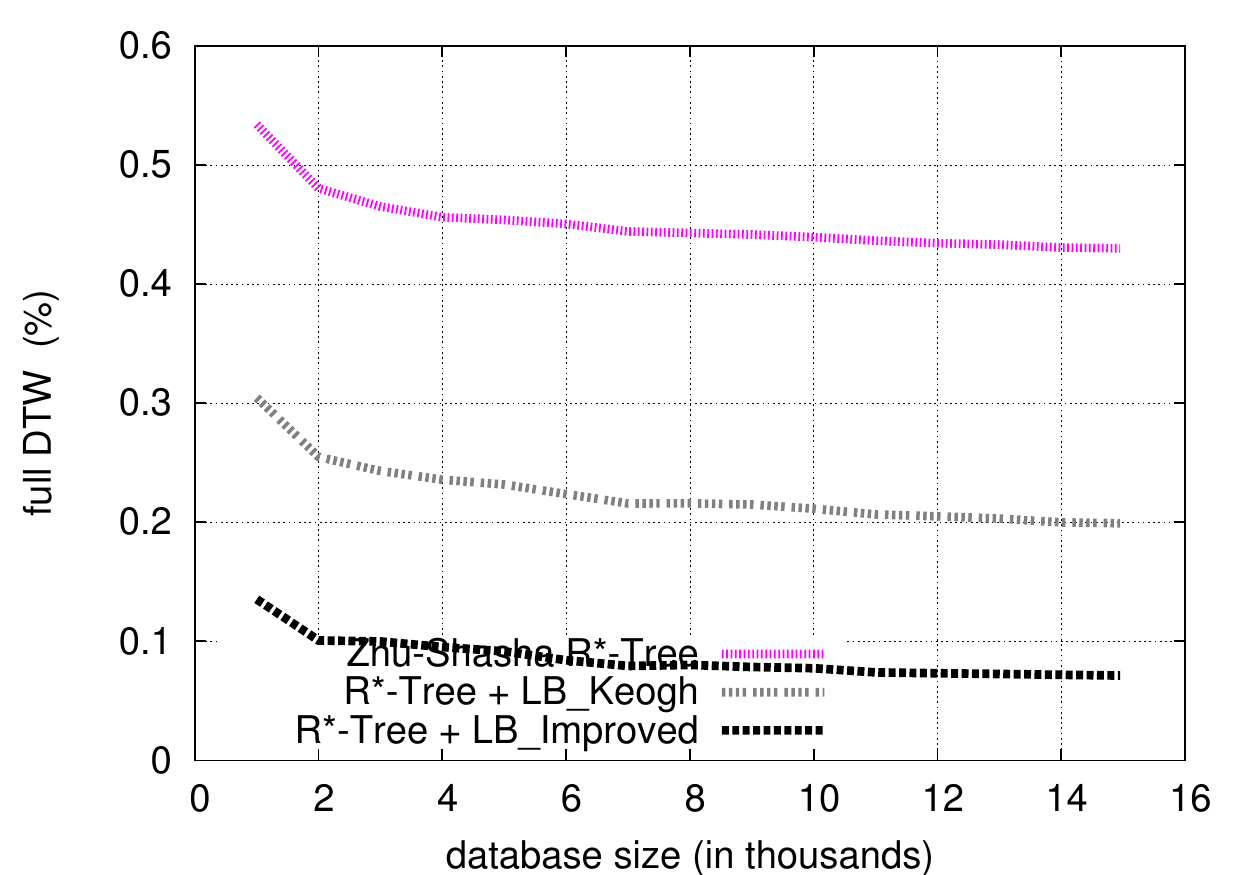}}  
\caption{Nearest-Neighbor Retrieval for the arrow-head shape data set\label{fig:perf5}}
\end{figure*}

 We also considered a large collection of time-series derived from shapes~\cite{keogh2006lse,keoghshapedata}.
The first data set is made of heterogeneous shapes which resulted in   5~844 1~024-sample times series.
The second data set is an arrow-head data set with  of 15~000 251-sample time series.
We extracted 50~time series 
from each data set, and we present the average nearest-neighbor retrieval times and pruning power
as we consider various fractions of each database (see Figs.~\ref{fig:perf4} and \ref{fig:perf5}).
The results are similar:  LB\_Improved has twice the pruning power than LB\_Keogh, \textsc{R*-tree+LB\_Improved}
is twice as fast as \textsc{R*-tree+LB\_Keogh} and over 3~times faster than 
the Zhu-Shasha R*-tree.

\subsection{Locality constraint}
  
\begin{figure*}
\centering
  \subfloat[ $w=5\%$]{\includegraphics[width=0.45\textwidth]{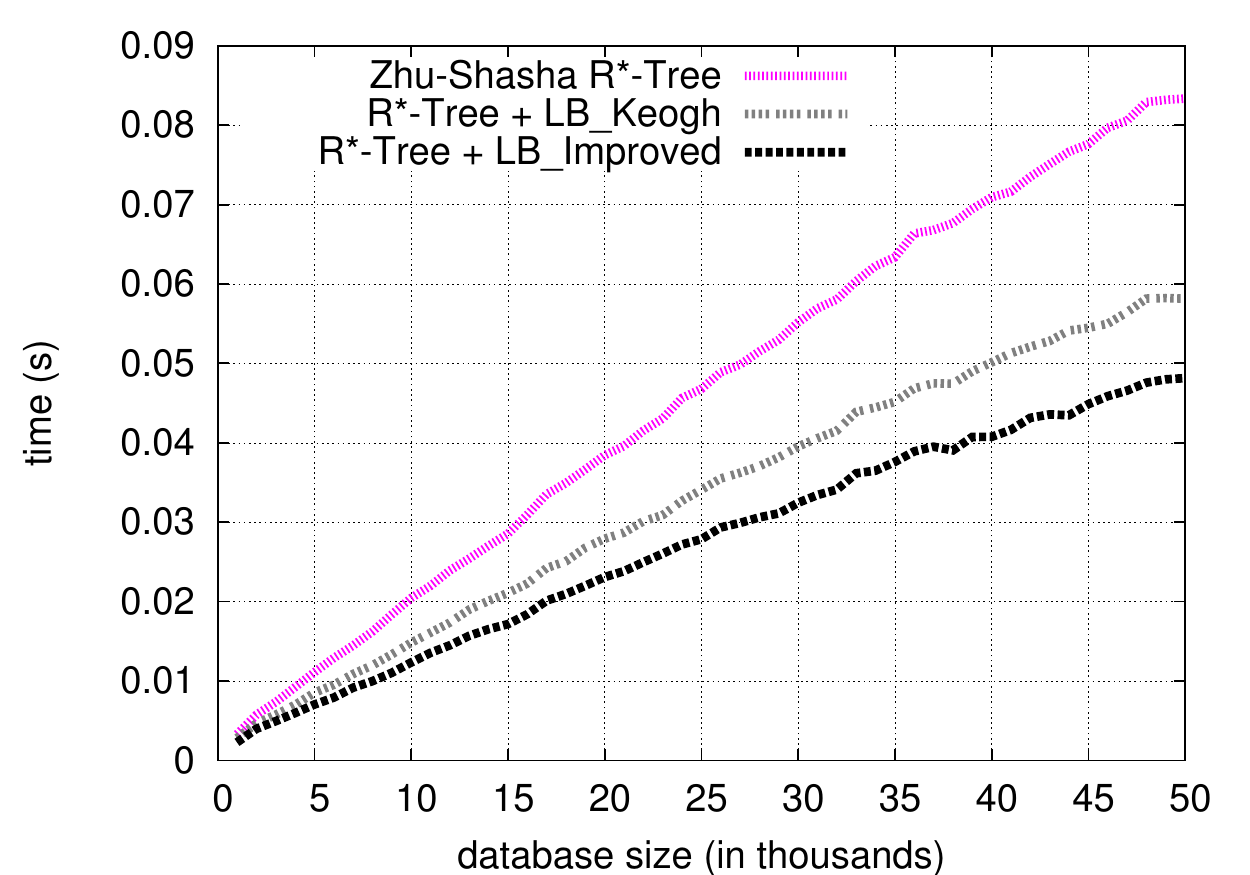}}
  \subfloat[ $w=20\%$]{\includegraphics[width=0.45\textwidth]{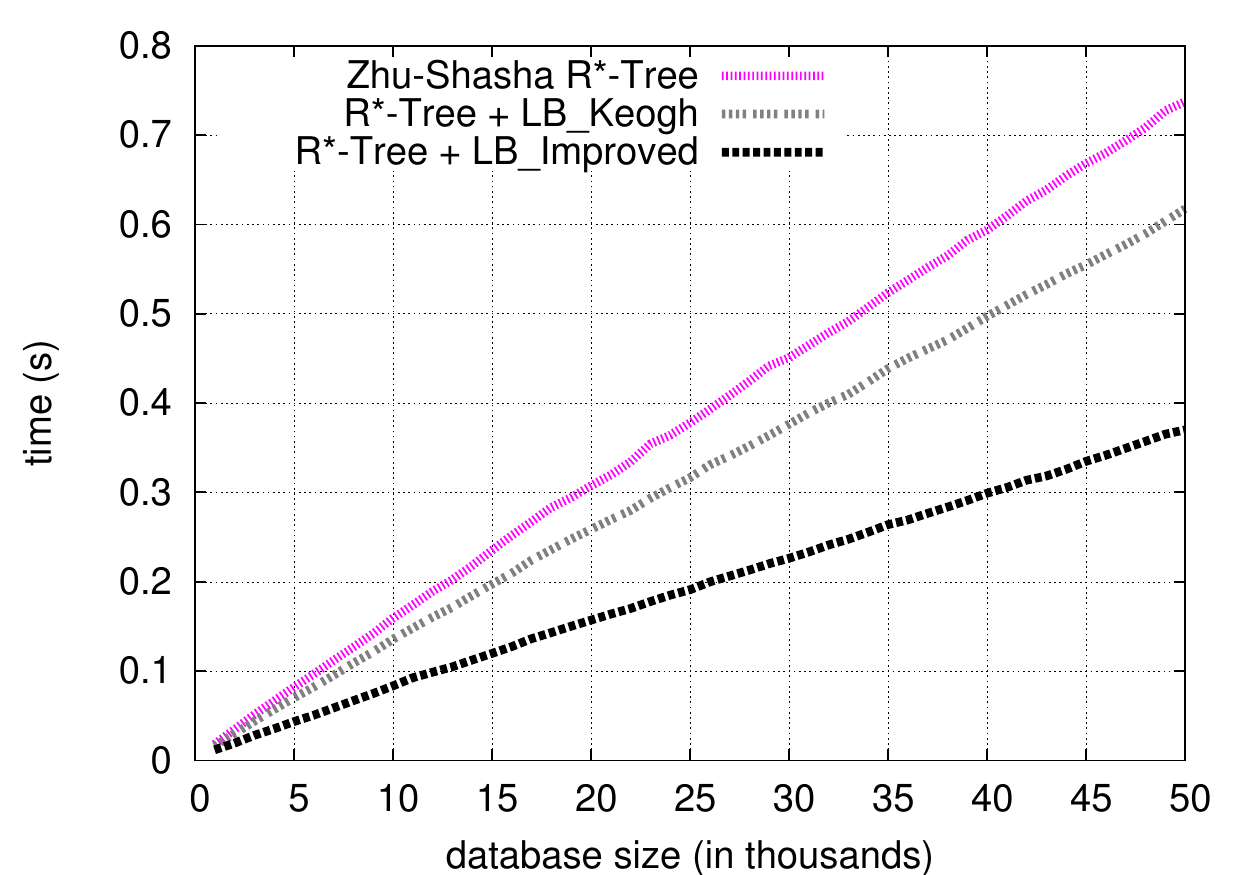}} 
\caption{Average Nearest-Neighbor Retrieval Time for the 256-sample random-walk data set\label{fig:variousw0}}
\end{figure*}

\begin{figure*}
\centering
  \subfloat[ $w=5\%$]{\includegraphics[width=0.45\textwidth]{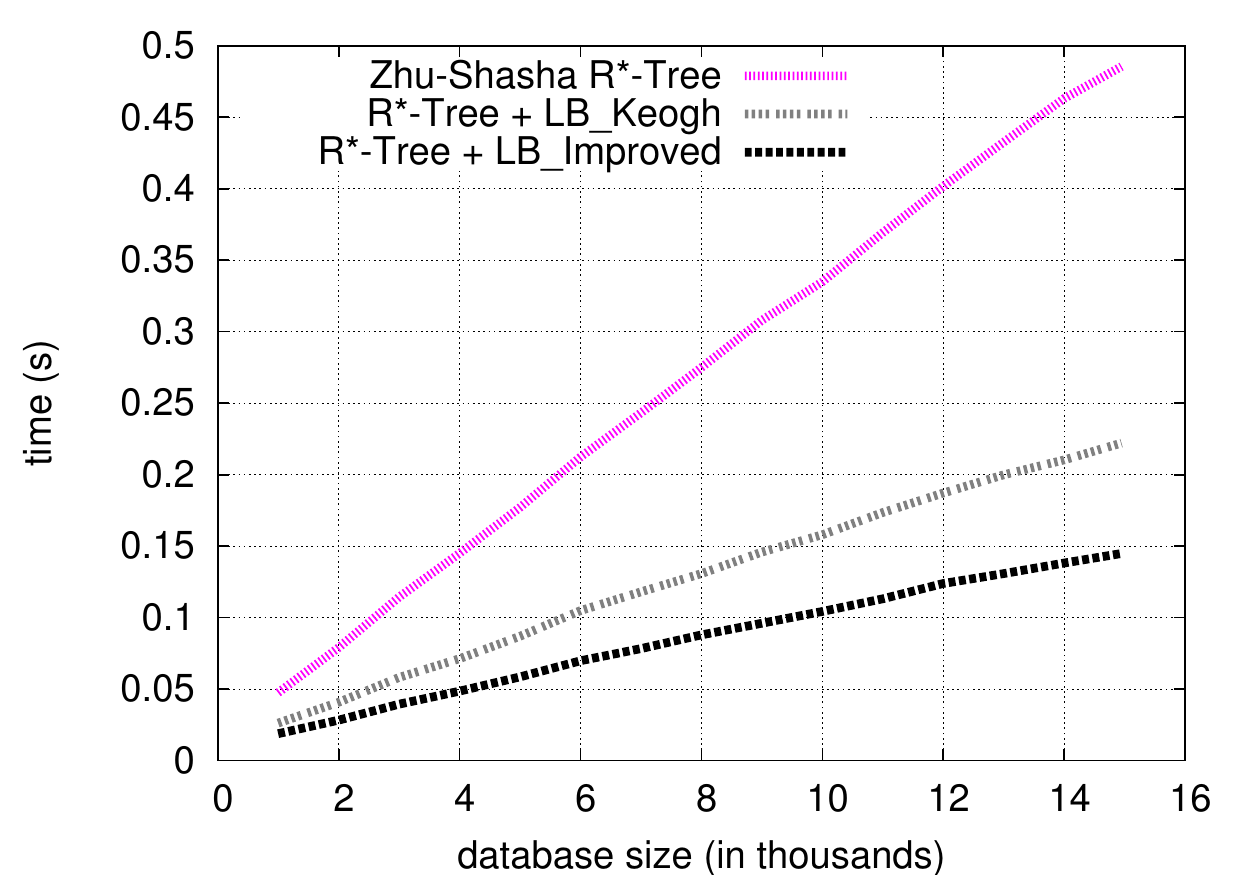}}
  \subfloat[$w=20\%$]{\includegraphics[width=0.45\textwidth]{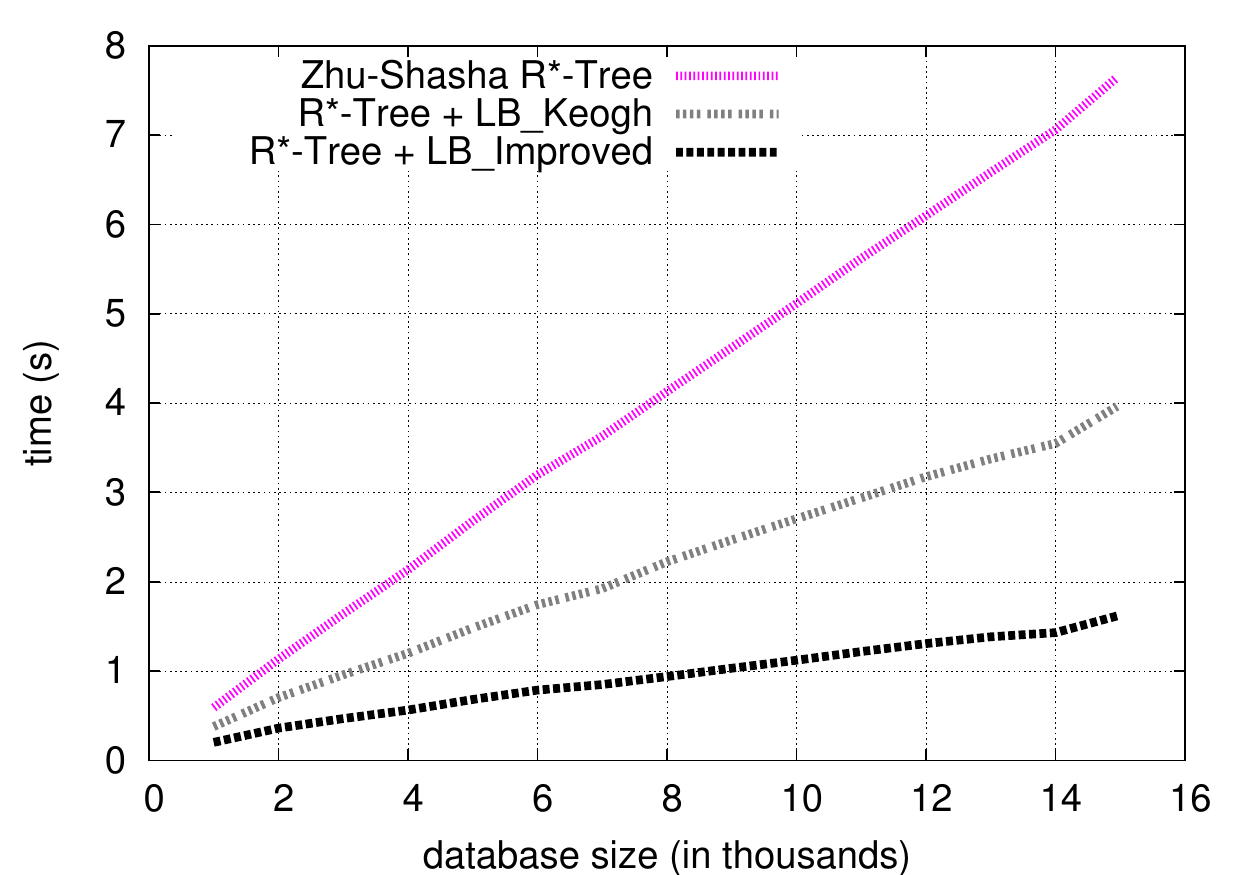}}  
\caption{Average Nearest-Neighbor Retrieval Time for the arrow-head shape data set\label{fig:variousw1}}
\end{figure*}

  The locality constraint has an effect on retrieval times: a large value of $w$ makes the problem
  more difficult and reduces the pruning power of all methods. In Figs.~\ref{fig:variousw0}
  and \ref{fig:variousw1}, we present the retrieval times for $w=5\%$ and $w=20\%$. 
  The  benefits of 
  \textsc{R*-tree}+\textsc{LB\_Improved} remain though they are  less significant for small
  locality constraints. Nevertheless, even in this case, \textsc{R*-tree}+\textsc{LB\_Improved}  can still be three times faster than Zhu-Shasha R*-tree. For all our data sets and for all values of $w\in \{5\%, 10\%, 20\%\}$,
  \textsc{R*-tree}+\textsc{LB\_Improved} was always at least as fast as the
  Zhu-Shasha R*-tree algorithm alone.
  
 \section{Conclusion}
 
 We have shown that a two-pass pruning technique can improve the retrieval speed
 by three times or more in several time-series databases.
 In our implementation, LB\_Improved required slightly more computation than LB\_Keogh,
 but its added pruning power was enough to make the overall computation
  several times faster.
Moreover, we showed that pruning candidates left from the
  Zhu-Shasha R*-tree with the full LB\_Keogh alone---without dimensionality reduction---was
  enough to significantly boost the speed and pruning power. 
  On some synthetic data sets, neither LB\_Keogh nor LB\_Improved
  were able to prune enough candidates, making all algorithms
  comparable in speed.
 

 \section*{Acknowledgements}
 The author is supported by NSERC 
grant 261437 and FQRNT grant 112381.

\bibliographystyle{elsart-num}
\bibliography{../maxminalgo,../lbkeogh}

\textbf{About the Author}--—DR. DANIEL LEMIRE received a B.Sc. and a M.Sc. in Mathematics from the University
of Toronto in 1994 and 1995. He received his Ph.D. in Engineering Mathematics from the Ecole Polytechnique and
the Universit\'e de Montr\'eal in 1998. He completed a post-doctoral fellowship at the Institut de g\'enie
biom\'edical and worked as consultant in industry. From 2002 to 2004, he was a research officer
at the National Research Council of Canada (NRC). He is now a professor at the Universit\'e du Qu\'ebec \`a Montr\'eal (UQAM)
where he teaches Computer Science. His research interests include data warehousing, OLAP and time series.

\appendix

\section{Some Properties of Dynamic Time Warping}
\label{sec:properties}

The DTW distance can be counterintuitive. 
As an example, if $x,y,z$ are three time series such that
 $x \leq y \leq z$ pointwise,
then it does not follow that $\text{DTW}_p(x,z) \geq \text{DTW}_p(z,y)$.
Indeed, choose $x=7,0,1,0$, $y=7,0,5,0$, and $z=7,7,7,0$,
then $\text{DTW}_\infty(z,y)=5$ and  $\text{DTW}_\infty(z,x)=1$.
Hence, we review some of the mathematical properties
of the DTW\@. 

The warping path aligns $x_i$ from time series $x$ and $y_j$ from time series $y$ if
$(i,j) \in \Gamma$. The next proposition is a general constraint on warping paths.

\begin{proposition}\label{prop:warpingpathprop} Consider any two time series $x$ and $y$. 
For any minimal warping path,  if $x_i$ is aligned with $y_j$, then either $x_i$ is aligned only with $y_j$
or $y_j$ is aligned only with $x_i$. Therefore the length of a minimal warping path is at most $2n-2$ when $n>1$.
\end{proposition}
\begin{pf} Suppose that the result is not true. Then there is $x_k, x_i$ and $y_l,y_j$ such
 that $x_k$ and $x_i$ are aligned with $y_j$, and $y_l$ and $y_j$ are aligned with $x_i$. We can delete $(k,j)$ from the warping path and  still have a warping path. A contradiction.
 
Next, we show that warping path is no longer than $2n-2$. Let $n_1$ be the number of points in $x$ aligned with only
one point in $y$, and let $n_2$ be the number of points in $y$ aligned with only one point in $x$.
The cardinality of a minimal warping path is bounded by $n_1+n_2$.
If $n_1=n$ or $n_2=n$, then $n_1=n_2=n$ and the warping path has cardinality $n$ which is no larger
than $2n-2$ for $n>1$. Otherwise,
$n_1\leq n-1$ and $n_2\leq n-1$, and $n_1+n_2<2n-2$.
\end{pf}

The next lemma shows that the DTW becomes the $l_p$ distance when either $x$ or $y$ is constant.

\begin{lemma}\label{lemma:yconstant}
For any $0 < p \leq \infty$, if $y=c$ is a constant, then $\text{NDTW}_p(x,y) = \text{DTW}_p(x,y)= \Vert x-y\Vert_p$.
\end{lemma}

When $p=\infty$, a stronger result is true: if $y=x+c$ for some constant $c$,
then $\text{NDTW}_\infty(x,y) = \text{DTW}_\infty(x,y)= \Vert x-y\Vert_\infty$. Indeed, 
 $\text{NDTW}_\infty(x,y)\geq \vert \max(y)-\max(x) \vert = c = \Vert x-y\Vert_\infty\geq \Vert x-y\Vert_\infty$ which
 shows the result.
This same result is not true for $p<\infty$: for $x=0,1,2$ and $y=1,2,3$, we have
$\Vert x-y\Vert_p=\sqrt[p]{3}$ whereas $\text{DTW}_p(x,y)=\sqrt[p]{2}$.
 However, the DTW is translation invariant:  
$\textrm{DTW}_p(x,z)=\textrm{DTW}_p(x+b,z+b)$
and $\textrm{NDTW}_p(x,z)=\textrm{NDTW}_p(x+b,z+b)$ for any scalar $b$ and $0<p\leq \infty$.

In classical analysis, we have that $n^{1/p-1/q} \Vert x\Vert_q \geq \Vert x \Vert_p$~\cite{folland84} for  $1\leq p < q \leq \infty$. A similar results is true for the DTW and it allows us to conclude that $\text{DTW}_p(x,y)$ and $\text{NDTW}_p(x,y)$ decrease monotonically as $p$ increases.

\begin{proposition} For $1\leq p < q \leq \infty$, we have that 
$(2n-2)^{1/p-1/q}  \text{DTW}_q(x,y) \geq \text{DTW}_p(x,y)$ 
where $n$ is the length of $x$ and $y$. The result also holds for the non-monotonic DTW.
\end{proposition}
\begin{pf}Assume $n>1$. The argument is the same for the monotonic or non-monotonic DTW\@.
 Given $x,y$ consider the two aligned (and extended) time series $x', y'$
 such that  $\text{DTW}_q(x,y)=\Vert x'-y' \Vert_q$. Let $n_{x'}$ be the length of
 $x'$ and $n_{y'}$ be the length of $y'$.
 As a consequence of  Proposition~\ref{prop:warpingpathprop},
 we have $n_{x'} =n_{y'} \leq 2n-2$. From classical analysis, we
 have $n_{x'} ^{1/p-1/q} \Vert x'-y'\Vert_q \geq \Vert  x'-y' \Vert_p$,
 hence $\vert 2n - 2\vert ^{1/p-1/q} \Vert x'-y'\Vert_q \geq \Vert  x'-y' \Vert_p$  
 or $\vert 2n - 2\vert ^{1/p-1/q} \text{DTW}_q(x,y) \geq \Vert  x'-y' \Vert_p$.
Since $x',y'$ represent a valid warping path of $x,y$,
then  $\Vert x'-y' \Vert_p \geq \text{DTW}_p(x,y)$ which concludes the proof. 
\end{pf}

\section{The Triangle Inequality}
\label{sec:ti}
The DTW is commonly used as a similarity measure:  $x$ and $y$ are similar
if $\text{DTW}_p(x,y)$ is small. Similarity measures
often define equivalence relations:  $A\sim A$ for all $A$ (reflexivity), $A\sim B \Rightarrow B \sim A$ (symmetry) and $A\sim B \land B \sim C \Rightarrow A \sim C$ (transitivity).

The DTW is  reflexive and symmetric, but it is not transitive. 
  Indeed, consider the following time series:
  \begin{align*}
  X&= \underbrace{0,0,\ldots,0,0}_{2m+1\text{ times}},\\
  Y& =\underbrace{0,0,\ldots,0,0}_{m\text{ times}}, \epsilon, \underbrace{0,0,\ldots,0,0}_{m\text{ times}},\\
  Z& =0, \underbrace{\epsilon,\epsilon,\ldots,\epsilon,\epsilon}_{2m-1\text{ times}}, 0.  
  \end{align*}
  We have that $\text{NDTW}_p(X,Y)=\text{DTW}_p(X,Y)=\vert \epsilon\vert $, $\text{NDTW}_p(Y,Z)=\text{DTW}_p(Y,Z)=0$,  $\text{NDTW}_p(X,Z)=\text{DTW}_p(X,Z)=\sqrt[p]{(2m-1) }\vert \epsilon\vert$ for $1\leq p<\infty$ and $w=m-1$.
  Hence, for $\epsilon$ small and $n\gg 1/\epsilon$, we have that  $X\sim Y$ and $Y\sim Z$, but $X \not\sim Z$. This example proves
  the following lemma.
 
 \begin{lemma}\label{lemma:ti}For $1\leq p < \infty$ and $w>0$, neither $\text{DTW}_p$ nor $\text{NDTW}_p$  satisfies
 a triangle inequality of the form $d(x,y)+d(y,z)\geq c d(x,z)$ where $c$ is independent of the length of the time series
 and of the locality constraint. 
 \end{lemma}
 
 This theoretical result is somewhat at odd with practical experience.
 Casacuberta et al. found no triangle inequality violation in about 15~million
  triplets of voice recordings~\cite{casacuberta1987mpd}. 
To determine whether we could expect violations of the triangle inequality in
practice, we ran the following experiment. We used 3~types of 100-sample time series: 
 white-noise times series 
defined by $x_i = N(0,1)$ where $N$ is the normal distribution,  random-walk 
time series defined by $x_i = x_{i-1}+N(0,1)$ and $x_1=0$, and the 
Cylinder-Bell-Funnel
time series proposed by Saito~\cite{921732}. For each type, we generated
100~000~triples of time series $x,y,z$ and we computed the histogram of the
function \begin{align*}C(x,y,z)=\frac{\text{DTW}_p(x,z)}{\text{DTW}_p(x,y) + \text{DTW}_p(y,z)} \end{align*} for $p=1$ and $p=2$.
The DTW is computed without time constraints. Over the white-noise and Cylinder-Bell-Funnel time
series, we failed
to find a single violation of the triangle inequality: a triple $x,y,z$ for
which $C(x,y,z) > 1$. 
However, for the random-walk time series,  we found that 20\% and 15\% of the triples violated the triangle inequality
for $\text{DTW}_1$ and $\text{DTW}_2$.



 
 

The DTW satisfies a weak triangle inequality as the next theorem shows.

\begin{theorem}
Given any 3~same-length time series $x,y,z$ and $1\leq p\leq \infty$, we have 
\begin{align*}\text{DTW}_p(x,y)+\text{DTW}_p(y,z) \geq \frac{\text{DTW}_p(x,z)}{\min(2w+1,n)^{1/p}}\end{align*}
where $w$ is the locality constraint. The result also holds for the non-monotonic DTW.
\end{theorem}
\begin{pf} Let $\Gamma$ and $\Gamma'$ be  minimal warping paths between $x$ and $y$
and between $y$ and $z$. Let $\Gamma''= \{(i,j,k) | (i,j) \in \Gamma \text{ and } (j,k) \in \Gamma'\}$.
Iterate through the tuples $ (i,j,k)$  in $\Gamma''$ and construct the same-length
time series $x'', y'', z''$ from $x_i$, $y_j$, and $z_k$. By the locality constraint any match $(i,j)\in \Gamma$
corresponds to at most $\min(2w+1,n)$~tuples of the form $(i,j,\cdot)\in \Gamma''$, and similarly for
any match $(j,k)\in \Gamma'$. 
Assume $1\leq p< \infty$. We
 have that $\Vert x''-y'' \Vert_p^p = \sum_{(i,j,k)\in \Gamma''} \vert x_i-y_j \vert^p \leq \min(2w+1,n)\text{DTW}_p(x,y)^p$
and $\Vert y''-z'' \Vert_p^p =\sum_{(i,j,k)\in \Gamma''} \vert y_j-z_k \vert^p \leq \min(2w+1,n)\text{DTW}_p(y,z)^p$.
By the triangle inequality in $l_p$, we have
\begin{align*}
 \min(2w+1,n)^{1/p}(\text{DTW}_p(x,y)+\text{DTW}_p(y,z)) & \geq \Vert x''-y'' \Vert_p+\Vert y''-z'' \Vert_p\\ 
                                                   &\geq  \Vert x''-z'' \Vert_p\geq \text{DTW}_p(x,z).
 \end{align*}
 For $p=\infty$, $\max_{(i,j,k)\in \Gamma''} \Vert x_i-y_j \Vert_p^p =\text{DTW}_{\infty}(x,y)^p$
 and
 $\max_{(i,j,k)\in \Gamma''} \vert y_j-z_k \vert^p = \text{DTW}_{\infty}(y,z)^p$, thus proving the result
 by the triangle inequality over $l_{\infty}$. 
The proof is the same for the non-monotonic DTW\@.\end{pf}

The constant $\min(2w+1,n)^{1/p}$ is tight. Consider the example with time series $X,Y,Z$ presented before
 Lemma~\ref{lemma:ti}. We have $\text{DTW}_p(X,Y) + \text{DTW}_p(Y,Z)=\vert \epsilon\vert $ 
 and $\text{DTW}_p(X,Z)= \sqrt[p]{(2w+1) }\vert \epsilon\vert $. Therefore, we have
 \begin{align*}\text{DTW}_p(X,Y)+\text{DTW}_p(Y,Z)= \frac{\text{DTW}_p(X,Z)}{\min(2w+1,n)^{1/p}}.\end{align*}

A consequence of this theorem is that $\text{DTW}_\infty$ satisfies the
traditional triangle inequality.
 
\begin{corollary}\label{coro:tilinfty}
The triangle inequality  $d(x,y)+ d(y,z)\geq d(x,z)$ holds
for $\text{DTW}_\infty$ and $\text{NDTW}_\infty$.
\end{corollary}

Hence the $\text{DTW}_\infty$ is a pseudometric: it is a metric over equivalence
classes defined by  $x\sim y$ if and only if $\text{DTW}_\infty(x,y)=0$.
When no locality constraint is enforced ($w\geq n$), $\text{DTW}_\infty$ is equivalent
to the discrete
Fr\'echet distance~\cite{eitermannila94}.

\section{Which is the Best Distance Measure?}
\label{sec:whichmeasure}
The DTW can be seen as the minimization of  the $l_p$ distance
under warping. Which $p$ should we choose?
Legrand et al. reported  best results
for chromosome classification using $\text{DTW}_1$~\cite{legrand2007ccu}
as opposed to using $\text{DTW}_2$. However, they did not quantify the
benefits of $\text{DTW}_1$.
Morse and Patel reported  similar results with both 
$\text{DTW}_1$ and $\text{DTW}_2$~\cite{morse2007eaa}.

While they do not consider the DTW, Aggarwal et al.~\cite{656414} 
argue
that out of the usual $l_p$ norms, only the $l_1$ norm, and to a lesser
extend the $l_2$ norm, express a qualitatively meaningful
distance when there are numerous dimensions. They even report on classification-accuracy experiments
where fractional $l_p$ distances such as $l_{0.1}$ and $l_{0.5}$ fare
better.  Fran\c{c}ois et al.~\cite{francois2007cfd} made the theoretical
result more precise showing that under uniformity assumptions, lesser
values of $p$ are always better. 

%







To compare $\text{DTW}_1$, $\text{DTW}_2$, $\text{DTW}_4$ and 
$\text{DTW}_{\infty}$, we considered four different synthetic time-series data sets:
Cylinder-Bell-Funnel~\cite{921732},
  Control Charts~\cite{pham1998ccp},
 Waveform~\cite{breiman1998car}, and
Wave+Noise~\cite{gonzalez2000tsc}. The time series in each data sets have
lengths 128, 60, 21, and 40.
 The Control Charts data set has
6~classes of time series whereas the other 3~data sets have 3~classes each.
For each data set, we generated various databases having
a different number of instances per class: between 1 and 9
inclusively for Cylinder-Bell-Funnel and Control Charts, and
between 1 and 99 for Waveform and Wave+Noise. For a given
data set and a given number of instances, 50~different databases
were generated. For each
database, we generated 500~new instances chosen from a random 
class and we found a nearest neighbor in the database using $\text{DTW}_p$
for $p=1,2,4,\infty$ and using a time constraint of $w=n/10$. When
the instance is of the same class as the nearest neighbor,
we considered that the classification was a success.

\begin{figure*}
\centering
  \subfloat[Cylinder-Bell-Funnel]{\includegraphics[width=0.45\textwidth]{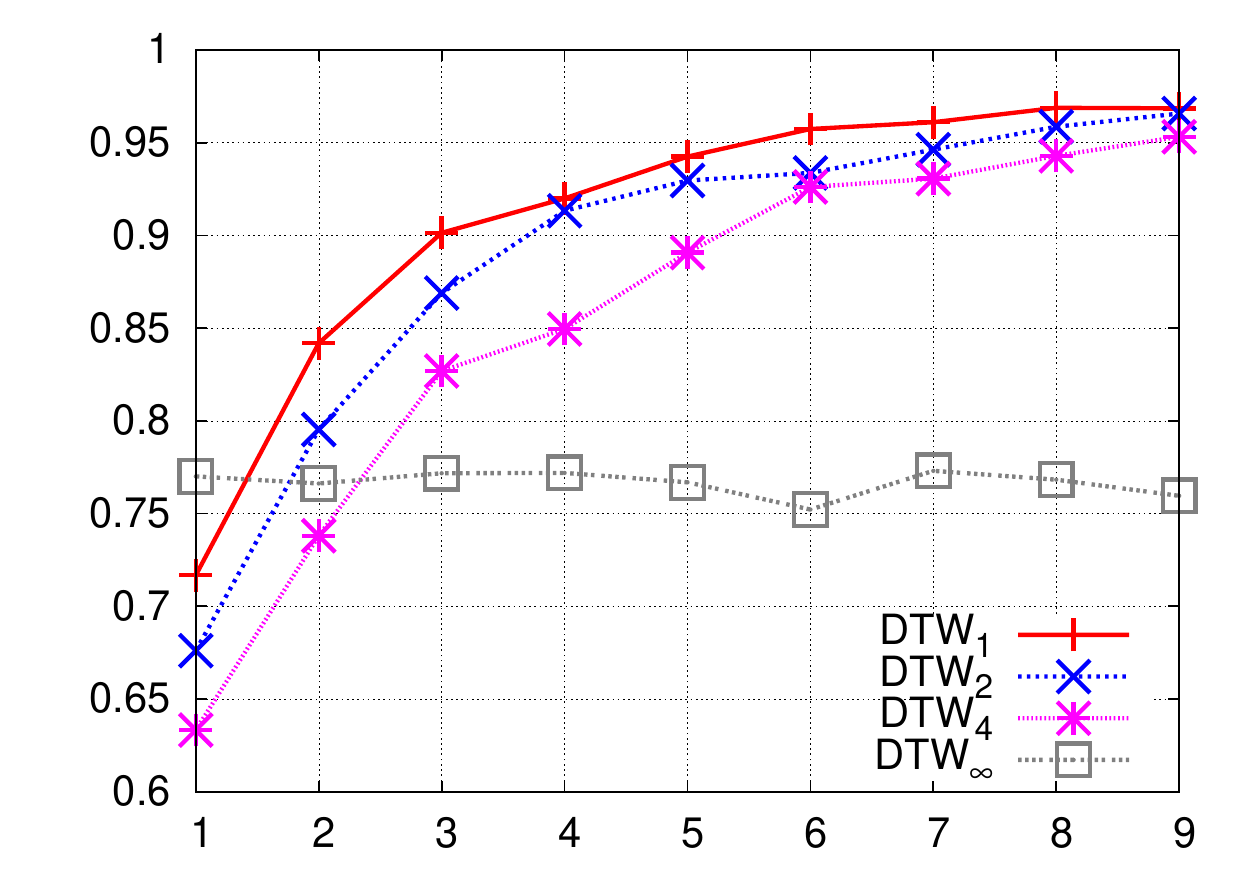}}
  \subfloat[Control Charts]{\includegraphics[width=0.45\textwidth]{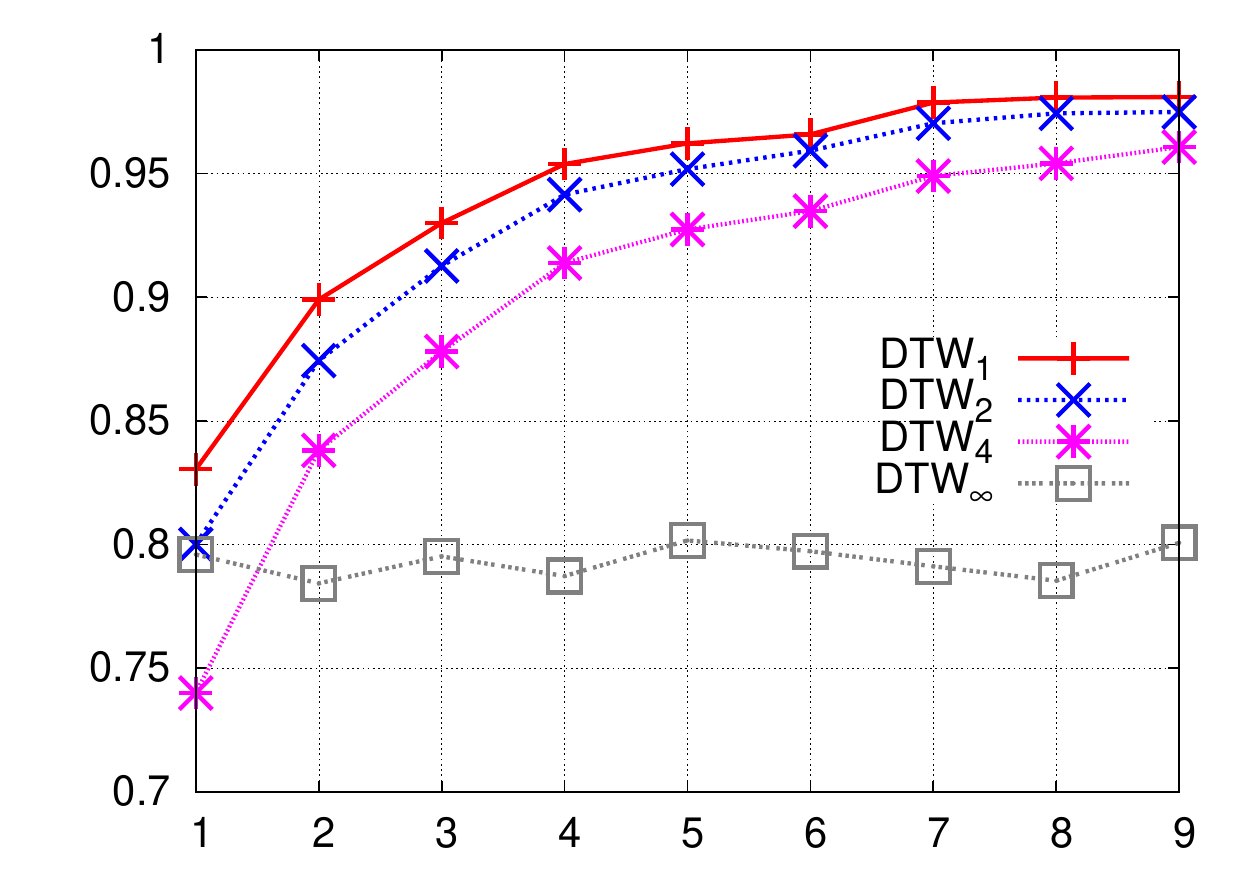}} \\
\subfloat[Waveform]{\includegraphics[width=0.45\textwidth]{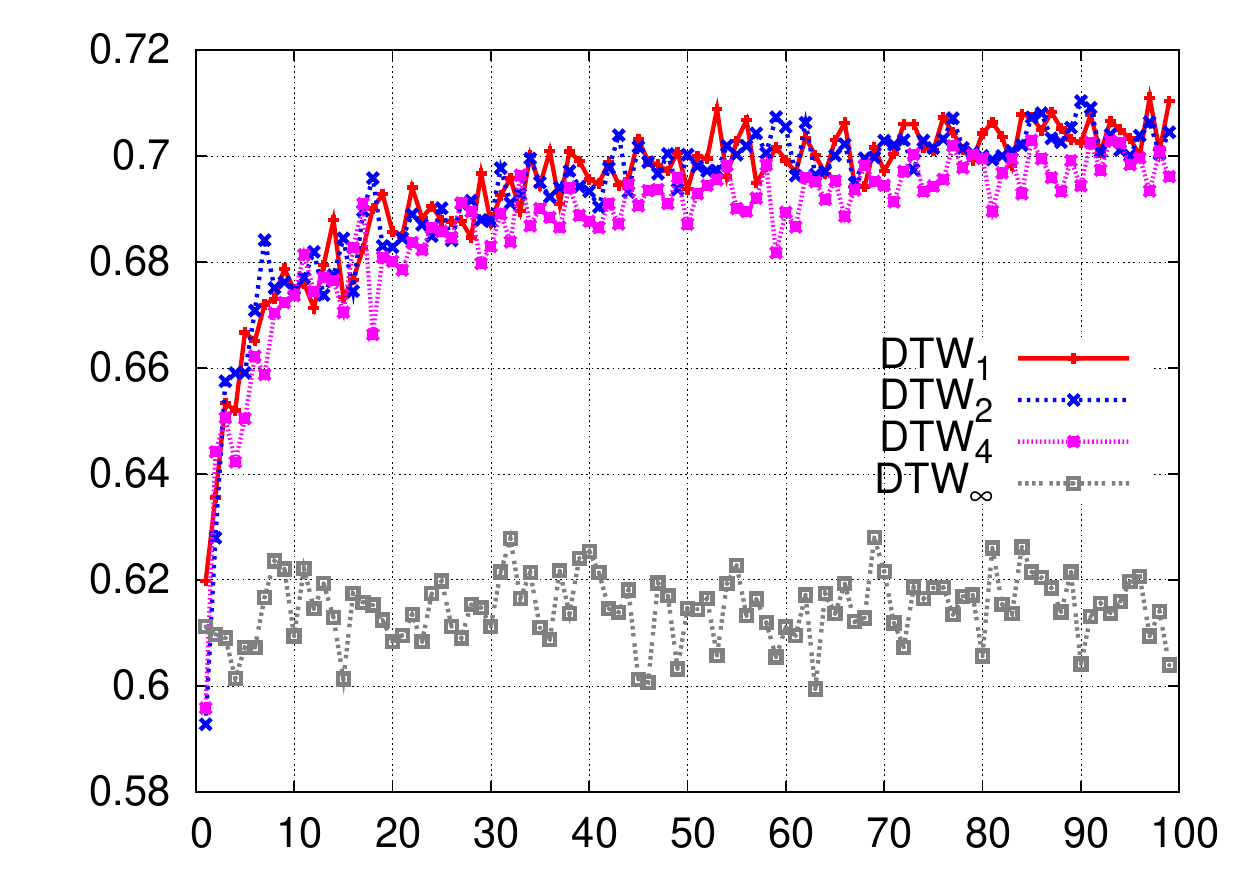}} 
  \subfloat[Wave+Noise]{\includegraphics[width=0.45\textwidth]{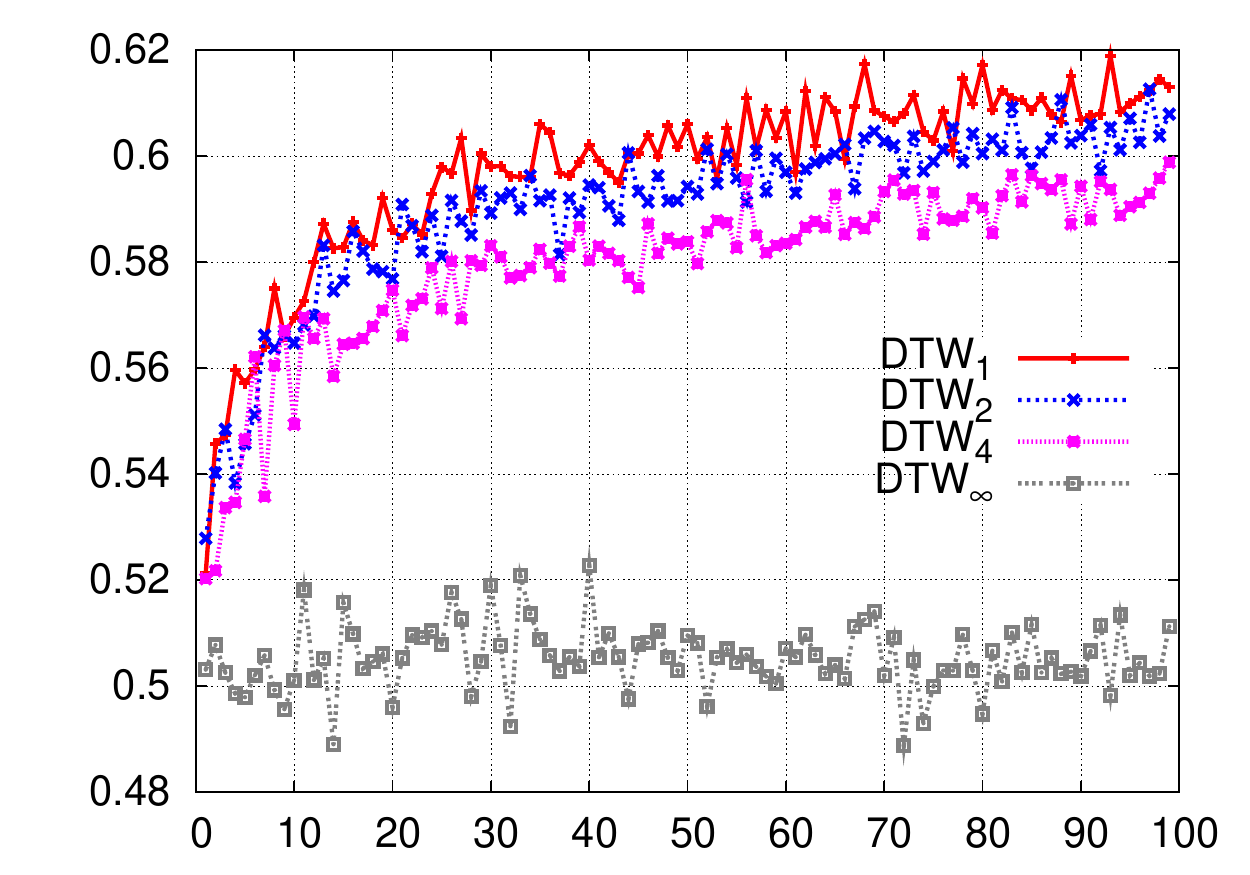}}
\caption{Classification accuracy versus the number of instances of each class in four data sets\label{fig:classaccuracy}}
\end{figure*}

The average
classification accuracies for the 4~data sets, and for various
number of instances per class is given in Fig.~\ref{fig:classaccuracy}.
The average is taken over 25~000~classification tests ($50\times 500$), over 50~different
databases.

Only when there are one or two instances of each class is $\text{DTW}_{\infty}$
competitive. Otherwise, the accuracy of the $\text{DTW}_{\infty}$-based classification
does not improve as we add more instances of each class. For the 
Waveform data set, $\text{DTW}_1$ and $\text{DTW}_2$ have comparable 
accuracies. 
For the other 3~data sets, 
$\text{DTW}_1$ has  a better nearest-neighbor classification
accuracy than $\text{DTW}_2$. 
Classification with $\text{DTW}_4$ has almost always a lower accuracy than
either $\text{DTW}_1$  or 
$\text{DTW}_2$.

Based on these results, $\text{DTW}_1$ is a good choice  to classify time series
whereas $\text{DTW}_2$ is a close second.  

\end{document}